\documentclass[]{sn-jnl}
\usepackage[utf8]{inputenc}
\usepackage{graphicx}%
\usepackage{multirow}%
\usepackage{amsmath,amssymb,amsfonts}%
\usepackage{amsthm}%
\usepackage{mathrsfs}%
\usepackage[title]{appendix}%
\usepackage{xcolor}
\usepackage{textcomp}%
\usepackage{manyfoot}%
\usepackage{booktabs}
\usepackage{listings}%
\usepackage{epsfig}
\begin{document}

\title{Coupled Nonlinear Schr{\"o}dinger (CNLS) Equations 
for two interacting electrostatic wavepackets
in a non-Maxwellian fluid plasma model}

\author*[1]{\fnm{N.} \sur{Lazarides}}\email{nikolaos.lazaridis@ku.ac.ae}
\author[1,2,3,4]{\fnm{Ioannis} \sur{Kourakis}}\email{ioanniskourakissci@gmail.com}

\affil*[1]{\orgdiv{Department of Mathematics}, 
\orgname{Khalifa University of Science and Technology}, 
\orgaddress{\city{Abu Dhabi}, \postcode{P.O. Box 127788},  
\country{United Arab Emirates}}}

\affil[2]{\orgdiv{Space and Planetary Science Center}, 
\orgname{Khalifa University of Science and Technology}, 
\orgaddress{\city{Abu Dhabi}, \postcode{P.O. Box 127788}, 
\country{United Arab Emirates}}}

\affil[3]{\orgdiv{Hellenic Space Center}, 
\orgaddress{\street{Leoforos Kifissias 178}, \city{Athens}, 
\postcode{GR-15231}, \country{Greece}}}

\affil[4]{\orgdiv{Department of Physics}, 
\orgname{National and Kapodistrian University of Athens}, 
\orgaddress{\street{Zografou}, \city{Athens}, \postcode{GR-15784}, 
\country{Greece}}}

\abstract{
The nonlinear dynamics of two co-propagating electrostatic wavepackets in a 
one-dimensional non-magnetized plasma fluid model is considered, from first 
principles. The coupled waves are characterized by different (carrier) 
wavenumbers and amplitudes. A plasma consisting of non-thermalized 
($\kappa-$distributed) electrons evolving against a cold (stationary) ion  
background is considered. The original model is reduced, by means of a 
multiple-scale perturbation method, to a pair of coupled nonlinear 
Schr{\"o}dinger (CNLS) equations for the dynamics of the wavepacket envelopes. 
For arbitrary wavenumbers, the resulting CNLS equations exhibit no known 
symmetry and thus intrinsically differ from the Manakov system, in general.

Exact analytical expressions have been derived for the dispersion, 
self-modulation (nonlinearity) and cross-modulation (coupling) coefficients 
involved in the CNLS equations, as functions of the wavenumbers ($k_1$, $k_2$) 
and of the spectral index $\kappa$ characterizing the electron profile. 
An analytical investigation has thus been carried out of the modulational 
instability (MI) properties of this pair of wavepackets, focusing on the role of 
the intrinsic (variable) parameters. Modulational instability is shown to occur 
in most parts of the parameter space. The instability window(s) and the 
corresponding growth rate are calculated numerically in a number of case studies. 
Two-wave interaction favors MI by extending its range of occurrence and by 
enhancing its growth rate. Growth rate patterns obtained for different $\kappa$ 
index (values) suggest that deviation from thermal (Maxwellian) equilibrium, for 
low $\kappa$ values, leads to enhances MI of the interacting wave pair.

To the best of our knowledge, the dynamics of two co-propagating wavepackets in 
a plasma described by a fluid model with $\kappa-$distributed electrons is 
investigated thoroughly with respect to their MI properties as a function of 
$\kappa$ for the first time, in the framework of an asymmetric CNLS system whose 
coefficients present no obvious symmetries for arbitrary $k_1$ and $k_2$.
Although we have focused on electrostatic wavepacket propagation in nonthermal 
(non-Maxwellian) plasma, the results of this study are generic and may be used 
as basis to model energy localization in nonlinear optics, in hydrodynamics or 
in dispersive media with Kerr-type nonlinearities where MI is relevant.
}

\maketitle
\section{Introduction}

Research on plasma dynamics has grown significantly during the last decades, 
because of the prospects for a wide range of applications related to space and 
fusion, among others. Plasmas in a state of collisional equilibrium exhibit a 
Maxwellian distribution. Space plasmas, however, are very often in stationary 
states out of equilibrium that can be described by kappa distributions, which 
are characterized by a Maxwellian-like ``core'' and a high-energy tail that 
follows a power law. Such distributions have been observed in the solar wind
and planetary magnetospheres 
\cite{Livadiotis2013,Nicolaou2018,Nicolaou2020,Livadiotis2019,Saberian2022}.
The intrinsic dispersion and nonlinearity in plasmas can be balanced to lead 
to collective dynamics and the emergence of localized excitation in the form 
of solitons, breathers, rogue waves, or shocks. Plasma fluid models 
mimicking Navier-Stokes equation in hydrodynamics 
(see e.g. in \cite{Kalita2023,Madhukalya2023}), 
can be reduced through standard perturbative methods to the celebrated 
nonlinear Schr{\"o}dinger (NLS) equation known to govern the envelope 
dynamics of a modulated wavepacket in nonlinear and dispersive media. In the 
context of plasma fluids, the NLS equations has been derived for the first 
times in the early '70s \cite{Shimizu1972,Kakutani1974}, and subsequently 
for a large variety of electrostatic plasma models.
Recently, its localized solutions (envelope solitons, breathers) have in fact 
been associated with freak or rogue waves in plasmas 
\cite{Kourakis2005,Chowdhury2018,Singh2020,Sarkar2020}. Formally analogous
NLS equations focusing on electromagnetic waves in plasmas modeled by 
fluid-Maxwell equations have been also derived
\cite{Borhanian2009,Borhanian2015,Veldes2013b}. In recent years, the kappa
distribution of electrons has been used more and more in NLS-reduced plasma
fluid models \cite{Borhanian2013,Sultana2015,Sultana2018,Kuldeep2019,Arham2021,
Sharmin2021,Yahia2021}.

Different versions of NLS equations have been also employed in other contexts,
e.g., in the investigation of the dynamics of blow-up solutions in trapped 
quantum gases in Gross-Pitaevskii equation \cite{Xie2018} which in a sense  
generalizes the three-dimensional NLS equation. Alternatively, plasma fluid 
model may be reduced to the Korteweg - de Vries (KdV) equation, which has been 
recently investigated in a generalized form (i.e., with high power nonlinearities) 
with respect to solitary, compacton, and periodic types of solutions \cite{Li2018}. 
Furthermore, 
the NLS equation with higher order dispersion and parabolic nonlinearity has 
been found to admit dark, bright, kink, as well as triangular solitary wave 
solutions \cite{Xie2021}. Also, several types of localized excitations have been
constructed for extended Kadomtsev–Petviashvili equations of the B type, such as
dromion- and lump-lattice structures as well as various folded solitary waves
\cite{Li2022,Li2023}.

The consideration of two co-propagating wavepackets in a particular plasma 
fluid and their mutual interaction, results in a pair of coupled NLS (CNLS) 
equations, whose coefficients generally depend on the carrier wavenumber(s) 
of the respective carrier waves. Electron plasma (Langmuir) waves and ion 
acoustic waves \cite{Spatschek1978,Som1979,Kofane2020}, amidst other studies 
that focused mostly on electromagnetic modes\cite{McKinstrie1989,McKinstrie1990,
Luther1990,Luther1992,Kourakis2005b,Singh2013,Borhanian2017}.
Beyond plasma science, general forms of CNLS equations have been 
investigated in recent years with respect to the existence of 
vector solitons, breathers, and rogue wave solutions, in contexts including 
higher-order coupled NLS systems \cite{Guo2021,Zhou2022}, coupled 
mixed-derivative NLS equations \cite{Xiang2022,Jin2023,Wu2023}, variable
coefficients CNLS equations \cite{Zhang2018nd}, non-autonomous CNLS equations
\cite{Yu2019,Patel2021,Yang2022}, systems involving four-wave mixing terms
\cite{Wang2021}, coupled cubic-quintic NLS equations \cite{Yan2020}, 
space-shifted CNLS equations \cite{Ren2022} and non-autonomous partially
non-local CNLS equations \cite{Wu2022}, among others.

More specifically, pairs of CNLS equations have been employed as the workhorse 
model for systems that belong to diverse areas of science such in nonlinear 
optics, hydrodynamics, and dispersive media, for the investigation of their 
dynamics and their modulational instability properties. Below we refer to some 
of these work in which formally equivalent forms of CNLS equations to those 
presented in Section 2.3 or 2.4 were employed.

Systems of CNLS equations have been obtained in Ref. \cite{Ablowitz2015} for the 
description of interacting nonlinear wave envelopes 
and formation of rogue waves in deep water. A condition for rogue wave formation 
was introduced which relates the angle of interaction with the group velocities 
of these waves. The CNLS system exhibit modulational instability (MI), which 
develops much faster (i.e., it has higher growth rates) than in each of the 
corresponding decoupled CNLS equations, and constitutes an important aspect of 
rogue wave dynamics. Further, CNLS equations were employed as a model describing 
the formation of rogue waves of the Peregrine breather type in water, to confirm 
corresponding experimental evidence. Indeed, in Ref. \cite{He2022}, it was 
demonstrated that nonlinear wave focusing resulting in strong localization on the 
water surface which originates from MI evolving from two counter-propagating 
water waves. The observed experimental results are in excellent agreement with 
the corresponding results obtained from the hydrodynamic CNLS equation model. 
In another relevant work \cite{Onorato2006}, a weakly nonlinear model consisting 
of a pair of CNLS equations was derived for two water wave systems known as 
crossing sea states in the context of ocean waves which propagate in two 
different directions. It was shown that when the first wave is modulationally 
unstable, then a second wave propagating in different direction results in an 
increase of the MI growth rates and enlargement of the instability region 
through their interaction.

Furthermore, a pair of CNLS equations which describe the interaction between the 
high-frequency Langmuir and low-frequency ion-acoustic waves in a plasma, and 
breather and rogue wave solutions have been investigated via the Darboux 
transformation \cite{Meng2015}. The co-propagation of two circularly polarized 
strong laser pulses in a magnetized plasma in a weakly relativistic regime and 
their MI properties were investigated in Ref. \cite{Borhanian2017} within the 
framework of two CNLS equations. Further works on dispersive media include the 
investigation of a CNLS equations model resulted from Maxwell's equations with 
nonlinear permittivity and permeability in a left-handed metamaterial 
\cite{Lazarides2005}, for which several types of vector solitons were identified, 
the investigation of the MI profile of the coupled plane-wave solutions in 
left-handed metamaterials using CNLS equations in Ref. \cite{Kourakis2005}, and 
the investigation of the coupling between backward- and forward-propagating 
waves, with the same group velocity, in a composite right- and left-handed 
nonlinear transmission line \cite{Veldes2013}.

Perhaps the largest amount of work on CNLS equations has been performed in optics, 
in which long ago those equations were shown to govern light propagation in 
isotropic Kerr materials \cite{Haelterman1994}. In that work, the existence of 
a novel class of vector solitary waves was revealed, which bifurcate from 
circularly polarized solitons. Also, the generation of vector dark-soliton 
trains by induced MI in a highly birefringent fiber was experimentally observed 
in Ref. \cite{Seve1999}, in excellent agreement with theoretical predictions from 
numerical simulations of the CNLS equations. More recent works include the 
emergence of temporal patterns such as bright solitons and rogue waves in optical 
fibers investigated theoretically via a special version of CNLS equations, i.e., 
the Manakov system, which are initiated by the MI process \cite{Frisquet2016}. 
The Manakov system has been also recently investigated with respect to the 
existence of novel, nondegenerate fundamental solitons corresponding to 
different wavenumbers \cite{Stalin2019}. More optical solitons have been 
obtained for the Manakov system of CNLS equations in Ref. \cite{Yilmaz2022}, 
while analytical one- and two-soliton solutions in birefringent fibers have been 
obtained through the Hirota bilinear method in Ref. \cite{Huang2022}.

In the present work, a pair of CNLS equations is derived from a plasma model 
comprising of a cold inertial ion fluid evolving against a thermalized electron 
background that follows a kappa distribution, using a standard multiple 
scale approach (the Newell method). The coefficients of the CNLS equations have 
been calculated as functions of the wavenumbers of the two interacting waves 
and the coefficients of a Mc Laurin expansion of the kappa distribution (which 
depend on the spectral index $\kappa$). For arbitrary wavenumbers, the 
coefficients of the CNLS equations do not exhibit any known symmetry and hence 
the system is rendered non-ingtegrable. A compatibility condition is derived
for that system through a detailed modulational instability analysis, that 
provides a polynomial relation (fourth order in the perturbation frequency)
between the perturbation frequency and wavenumber \cite{Kourakis2006}. 
The MI properties of the CNLS equations are then investigated with respect to 
the wavenumbers of the two carrier waves and the wavenumber of the perturbation 
for several values of the spectral index $\kappa$ of the electron distribution.

In comparison with earlier similar works, which in the context of plasma are 
rather scarce, we should stress that the obtained CNLS equations for the envelops 
of the two interacting wavepackets are the most general as far as the values of 
their coefficient are concerned; those coefficients are derived exactly in terms 
of $k_1$, $k_2$, and $\kappa$ without any other additional assumption (e.g., 
equal group velocities of the wavepackets). The choice of the fluid model was 
made on the basis on simplicity; it contains only one parameter, the spectral 
index $\kappa$ (besides the externally determined $k_1$ and $k_2$). 
This constitutes advantage to other models since it isolates the effect of 
$\kappa$ on the investigated MI properties. Moreover, a thorough MI analysis is 
performed as a function of $\kappa$ for the first time, which illuminates its role 
for values that are relevant to space plasmas and have been deduced from 
observational data. A comparison of the results for small values of $\kappa$ 
(in the interval [2,3]) 
with those using the common Maxwell-Boltzmann electron distribution reveals 
dramatic differences as far as the strength of the growth rate of MI and the MI 
window are concerned, which have not been shown before.

In the following Section \ref{section2} the plasma fluid model equations are 
introduced, from which the CNLS equations are derived using the multiple-scale 
perturbation method. In Section \ref{section3}, we undertake an analytical study 
of the modulational instability analysis of plane wave solutions of the CNLS 
equations. A detailed parametric analysis of MI occurrence 
is carried out, based on numerical calculation of the growth rate in Section 
\ref{section4}. Our main results are briefly discussed and summarized in Section 
\ref{section5}.

\section{Derivation of a system of CNLS equations for two interacting wavepackets 
\label{section2}
}

\subsection{Plasma fluid model equations and electron distribution}

A one-dimensional (1D), non-magnetized plasma model is considered here for 
electrostatic excitations, that consists of a cold inertial ion fluid of number
density $n_i =n$ and velocity $u_i =u$, evolving against a thermalized, 
kappa-distributed electron background of number density $n_e$. The continuity, 
momentum and Poisson partial differential equation(s) governing the dynamics of 
that model are given by \cite{Lazarides2023}
\begin{eqnarray}
\label{eq01}
   \frac{\partial n}{\partial t} +\frac{\partial (n u)}{\partial x} =0, 
   \\
\label{eq02}
   \frac{\partial u}{\partial t} +u \frac{\partial u}{\partial x} =
   -\frac{\partial \phi}{\partial x}, 
   \\
\label{eq03}
   \frac{\partial^2 \phi}{\partial x^2} = n_e -n,
\end{eqnarray}
respectively, with $\phi$ being the electrostatic potential. In Eqs.
(\ref{eq01})-(\ref{eq03}) above, the spatial and temporal variables $x$ and $t$,
are normalized to the inverse ion plasma frequency $\omega_{pi}^{-1}$ and to the 
Debye screening length $\lambda_D$, respectively, where 
\begin{eqnarray}
\label{eq04}
   \lambda_D =\left(\frac{\varepsilon_0 k_B T_e}{z_i n_{i0} e^2}\right)^{1/2}, 
  \\
 \label{eq05} 
   \omega_{pi}=\left[\frac{n_{0, i} (z_i e)^2}{\varepsilon_0 m_i}\right]^{1/2} \, .
\end{eqnarray}
The number density variable(s) of the ions $n$ and the electrons $n_e$ are 
normalized to their respective equilibrium values $n_{i, 0}$ and 
$n_{e, 0} = z_i n_{i, 0}$, the ion velocity $u$ is normalized to 
\begin{equation}
\label{eq06}
   c_s = \lambda_D \, \omega_{pi} =\left(\frac{z_i k_B T_e}{m_e}\right)^{1/2},
\end{equation}
i.e. essentially the (ion-acoustic) sound speed. Finally, the electrostatic 
potential $\phi$ is normalized to $k_B T_e/e$. Adopting a standard  notation, 
the symbols $\varepsilon_0$, $k_B$, $T_e$, $e$, $m_i$, $z_i$ appearing in the 
latter expressions denote respectively the dielectric permittivity in vacuum, 
the Boltzmann's constant, the (absolute) electron temperature, the electron 
charge, the ion mass, and the degree of ionization ($z_i = q_i/e$, where $q_i$ 
is the ion charge). 

As announced above, we shall assume the electrons to 
follow a kappa-distribution 
\cite{Livadiotis2013,Pierrard2010,Livadiotis2017} that is given by 
\cite{Baluku2010,Hellberg2009} 
\begin{equation}
\label{eq07}
   n_e =\left( 1 -\frac{\phi}{\kappa -\frac{3}{2}} \right)^{-\kappa +\frac{1}{2} } .
\end{equation}
Note that this function is characterized by a real parameter  $\kappa$ 
(known as the spectral index). 

The kappa distribution adopted here for the nonthermal electrons is a 
straightforward replacement for the Maxwell-Boltzmann distribution when dealing 
with systems in stationary states out of equilibrium \cite{Livadiotis2009}. 
Such stationary states are commonly found in space and astrophysical plasmas 
\cite{Nicolaou2018,Nicolaou2020}, whose nature allows for 
extracting the spectral index $\kappa$ in each particular case. 
Possible values of $\kappa$ in Eq. (\ref{eq07}) lie in the interval $(3/2, \infty)$,
while for $\kappa =+\infty$ the kappa distribution reverts to the Maxwell-Boltzmann
one. 
Lower values of $\kappa$ are associated with a long tail in the particle (velocity) 
distribution, in account of a strong suprathermal component (i.e., for large velocity) 
\cite{Livadiotis2017}. This is a common occurrence e.g. in Space plasmas 
\cite{Livadiotis2013}.     

Space plasma observations have shown an empirical separation for the spectrum 
of the values of $\kappa$ into a near- ($3/2 < \kappa < 2.5$) and a far- 
($2.5 < \kappa < \infty$) equilibrium region. In particular, kappa 
distributions with $2 < \kappa < 6$
have been found to fit the observations and satellite data in the solar wind
\cite{Pierrard2010} (and references therein). In later sections, we are
using values of $\kappa$ evenly distributed in the interval $2 < \kappa < 6$ as
well as a large value of $\kappa =100$ for which practically the electron 
component in the considered plasma model has reached equilibrium 
(Maxwell-Boltzmann).

Kappa distributions have by now become one of the most widely used tools for 
characterizing and describing space plasmas and are known to affect the 
fundamental properties of plasma waves and also of electrostatic solitary 
waves \cite{Hellberg2012}.
For low values of the electrostatic potential $\phi$, i.e. for $|\phi| \ll  1$ (in 
normalized units), the kappa distribution Eq. (\ref{eq06}) can be expanded in a 
Mc Laurin series in powers of $\phi$ as
\begin{equation}
\label{eq08}
   n_e \approx  1 +c_1 \phi +c_2 \phi^2 +c_3 \phi^3 +\cdots \, ,
\end{equation}
from which we shall keep terms up to order $\phi^3$. In that case, the first three
expansion coefficients are given by
\begin{eqnarray}
\label{eq09}
   c_1 &=&\frac{\kappa -\frac{1}{2}}{\kappa -\frac{3}{2}} \,  ,  
   \\
\label{eq10}
   c_2 &=&\frac{\left(\kappa -\frac{1}{2}\right) \left(\kappa +\frac{1}{2}\right)}
             {2! \left(\kappa -\frac{3}{2} \right)^2} \,  , 
\\
\label{eq11}
   c_3 &=&\frac{\left(\kappa -\frac{1}{2}\right) \left(\kappa 
       +\frac{1}{2}\right) \left(\kappa +\frac{3}{2}\right)}
             {3! \left(\kappa -\frac{3}{2}\right)^3} \,  .
\end{eqnarray}

\subsection{Application of the reductive perturbation method}
We introduce fast and slow variables so that the derivatives are approximated as
\begin{eqnarray}
\label{eq120}
   \frac{\partial}{\partial t} \rightarrow 
    \sum_{k=0}^\infty \varepsilon^k \frac{\partial}{\partial t_k},
    ~~~
   \frac{\partial}{\partial x} \rightarrow  
   \sum_{k=0}^\infty \varepsilon^k \frac{\partial}{\partial x_k} 
\end{eqnarray}
where $\varepsilon$ is a small quantity, and $x_k=\varepsilon^k x$, 
$t_k=\varepsilon^k t$ (for $k = 0, 1, 2, 3, ...$). The dependent variables $n$, 
$u$, and $\phi$ can be expanded around their equilibrium value as
\begin{eqnarray}
\label{eq13}
   n=1 +\sum_{k=1}^\infty \varepsilon^k n_k, \\
\label{eq14}
   u=\sum_{k=1}^\infty \varepsilon^k u_k, \\
\label{eq15}
   \phi=\sum_{k=1}^\infty \varepsilon^k \phi_k.
\end{eqnarray}
The expansions of $\{n, u, \phi\}$ in Eqs. (\ref{eq13})-(\ref{eq15}) along 
with those for the derivatives in Eq. (\ref{eq120}) and the expansion of the 
electron distribution in Eq. (\ref{eq07}) are substituted into the model 
equations Eqs. (\ref{eq01})-(\ref{eq03}), in order to obtain sets of perturbative
equation of different orders in $\varepsilon$ up to order $\varepsilon^3$.

The fluid equations obtained at first order ($\propto \varepsilon^1$) read 
\begin{eqnarray}
\label{eq_i01}
   \frac{\partial n_1}{\partial t_0} +\frac{\partial (u_1)}{\partial x_0} =0,  \\
\label{eq_i02}
   \frac{\partial u_1}{\partial t_0} +\frac{\partial \phi_1}{\partial x_0} =0,  \\
\label{eq_i03}
   \frac{\partial^2 \phi_1}{\partial x_0^2} +n_1 -c_1 \phi_1 =0,
\end{eqnarray}
where $x_0 =x$ and $t_0 =t$. In these equations, the following {\em Ansatz} will be 
introduced
 \begin{eqnarray}
\label{eq_i04}
  S_1 =S_{1,1}^{(1)} \, e^{+i \theta_1} +S_{1,2}^{(1)} \, e^{+i \theta_2} +c.c.,
\end{eqnarray}
where $S_1=\{n_1, u_1, \phi_1\}$ and the phases are $\theta_j =k_j x -\omega_j t$  with 
$k_j$ and $\omega_j$ being the wavevector and the corresponding (angular) frequency of the 
$j-$th wave (for $j=1, 2$). Note that the derivatives with respect to $t=t_0$ and $x=x_0$ that 
appear in Eqs. (\ref{eq_i01})-(\ref{eq_i03}) do not act on the amplitude(s) 
$S_{1,j}^{(1)}$ --  -- (but only on the harmonic carrier functions, i.e. the 
exponentials, above), since the latter does not depend on these variables. After 
some straightforward calculations we get the equations ($j=1,2$)
\begin{eqnarray}
   -\omega_j n_{1,j}^{(1)} +k_j u_{1,j}^{(1)} =0, 
\nonumber \\
\label{eq_i05}
   -\omega_j u_{1,j}^{(1)} +k_j \phi_{1,j}^{(1)} =0,
 \\
  +n_{1,j}^{(1)} -(k_j^2 +c_1) \phi_{1,j}^{(1)} =0.
\nonumber
\end{eqnarray}
The equations for $j=1$ are not coupled to those with $j=2$ and thus the two 
waves do not interact at this order of approximation. As a result, two uncoupled 
$3\times 3$ systems with unknowns 
$\{n_{1,j}^{(1)}, u_{1,j}^{(1)}, \phi_{1,j}^{(1)}\}$ for $j=1$ and $j=2$,
respectively, are obtained as ($j=1,2$)
\begin{eqnarray}
\label{eq_i06}
\begin{bmatrix}
   -i \omega_j & +i k_j      &  0  \\
    0          & -i \omega_j & +i k_j  \\
    1          & 0           & -(k_j^2 +c_1) 
\end{bmatrix}
\begin{bmatrix}
   n_{1,j}^{(1)}     \\
   u_{1,j}^{(1)}     \\
  \phi_{1,j}^{(1)}       
\end{bmatrix}
 =
\begin{bmatrix}
  0    \\
  0    \\
  0     
\end{bmatrix}
,
\end{eqnarray}
For these systems to have a non-trivial solution, it is required that their 
determinants should vanish, i.e., 
\begin{eqnarray}
\label{eq_i07}
D_j =
\begin{vmatrix}
   -\omega_j & +k_j       &  0  \\
     0       & -\omega_1  & +k_j  \\
     1       & 0          & -(k_j^2 +c_1) 
\end{vmatrix}
=0,
\end{eqnarray}
which provides the linear frequency dispersion relation ($\omega$) and the 
corresponding group ($v_g$) and phase ($v_{ph}$)  velocity for the $j-$th 
wave ($j=1$ or 2)  as
\begin{eqnarray}
\label{eq_i08}
   \omega_j &=&\frac{k_j}{[k_j^2 +c_1(\kappa)]^{1/2}}, 
   \\
\label{eq_i08.2} 
   v_{g,j} &=& \frac{c_1(\kappa)}{[k_j^2 +c_1(\kappa)]^{3/2}},
   \\
\label{eq_i08.3} 
   v_{ph,j} &=& \frac{\omega_j}{k_j} =\frac{1}{[k_j^2 +c_1(\kappa)]^{1/2}}, 
\end{eqnarray}
where the dependence of the expansion coefficient $c_1$ on the spectral 
index $\kappa$ is explicitly indicated.
The systems of Eqs. (\ref{eq_i06}) can be solved in terms of the variables
($j=1,2$)
\begin{equation}
\label{eq_i09.2}
   \phi_{1,j}^{(1)} =\Psi_j
\end{equation}
to give
\begin{equation}
\label{eq_i09}
    n_{1,j}^{(1)} =\left(\frac{k_j}{\omega_j}\right)^2 \Psi_j, 
    ~~~~
    u_{1,j}^{(1)} =\frac{k_j}{\omega_j} \Psi_j.
\end{equation}
\begin{figure}[htp]
    \centering
    \includegraphics[width=12cm]{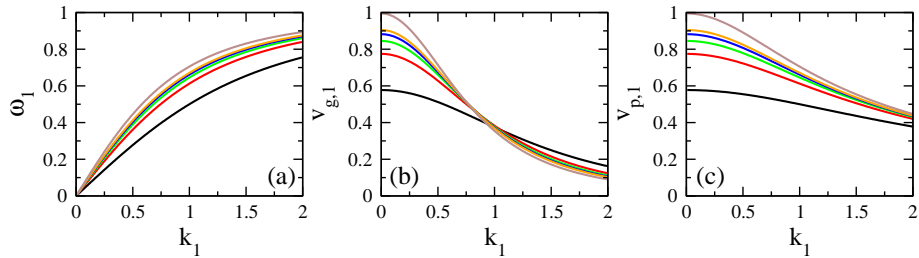}
    \caption{
    The frequency dispersion $\omega_1$ (a), 
    the corresponding group velocity $v_{g,1}$ (b) 
    as calculated from Eq. (\ref{eq_i08.2}), 
    and the phase velocity $v_{ph,1} =\omega_1/k_1$ (c)
    as calculated from Eq. (\ref{eq_i08.3}),
    as a function of the wavenumber $k_1$ for $\kappa=2$ (black), 
    $\kappa=3$ (red), $\kappa=4$ (green),
    $\kappa=5$ (blue), $\kappa=6$ (orange), $\kappa=100$ (brown).
    }
    \label{fig2}
\end{figure}
The frequency dispersion $\omega_1$ and the group velocity $v_{g,1}$ which have 
been calculated from Eqs. (\ref{eq_i08}) and (\ref{eq_i08.2}) are shown in Fig. 
\ref{fig2} as a function of the wavenumber $k_1$ for six values of the spectral 
index $\kappa$. The frequency dispersion curves seem to have the same qualitative 
features for any value of $\kappa$. However, the frequency (or fixed $k_1$) 
increases with increasing $\kappa$ for any $k_1$ in the (physically relevant) 
interval shown, while it saturates for large spectral index $\kappa$ to its 
Maxwell-Boltzmann value. Note that the curve obtained for $\kappa =100$, shown 
as brown curve in Fig. \ref{fig2}(a), is practically identical to a curve that 
would have been obtained using the Maxwell-Boltzmann distribution for the 
electrons.

The corresponding group velocity curves in Fig. \ref{fig2}(b), exhibit two 
distinct behaviors; for low $k_1$ (i.e., $k_1 \lesssim 0.9$) the group velocity 
$v_{g,1}$ increases with increasing $\kappa$ just as the frequency does in 
Fig. \ref{fig2}(a). However, for high $k_1$ (i.e., $k_1 \gtrsim 0.9$) that
tendency is reversed, and the group velocity $v_{g,1}$ decreases with increasing 
$\kappa$. Note that for very low $k_1$ (i.e., for $k_1 \rightarrow 0$), the 
group velocity increases from $v_{g,1} \simeq 0.59$ for $\kappa =2$ to 
$v_{g,1} =1$ for $\kappa =100$ (practically the same as in the case of 
Maxwell-Boltzmann distributed electrons, as expected for large values of $\kappa$). 
This constitutes a relative increase of up to $\simeq 70\%$ in group velocity 
(value) between the extreme values of $\kappa$ adopted here. We draw the conclusion 
that lower values of $\kappa$, i.e. a stronger deviation from the thermal 
(Maxwellian) picture result in lower group velocity (i.e. slower wavepackets/envelopes) 
for electrostatic wavepackets of this kind, at least for long wavelengths 
(i.e. $\gtrsim$ the Debye length, more or less). The inverse trend is observed 
for short (sub-screening-radius) wavelengths (i.e. for wavenumber exceeding unity, 
roughly), since  lower $\kappa$ values entail a slightly faster $v_g$ (value).

The phase velocity $v_{ph, 1}$, on the other hand, 
exhibits a simpler behavior, as illustrated in Fig. \ref{fig2}(c). 
It turns out that $v_{ph, 1}$ increases with increasing $\kappa$ 
for all wavenumbers, while at the same time the phase speed decreases asymptotically 
towards zero for large $k_1$ (i.e. for vanishing wavelength), regardless of the value 
of $\kappa$, as expected from the analytical expression  (\ref{eq_i08.3}). 
For negligible $k_1$ (i.e. in the infinite wavelength limit), the expressions for 
both group and phase speeds recover the sound speed 
$\lim_{k_1 \rightarrow 0} v_{g,1} =\lim_{k_1 \rightarrow 0} v_{ph,1} = 1/\sqrt{c_1(\kappa)}$,
as can be inferred from Eqs. (\ref{eq_i08.2}) and (\ref{eq_i08.3}). 

Obviously, all of the observations in the latter two paragraphs hold for the second wave 
also, upon a simple subscript permutation $1 \rightarrow 2$.

The equations obtained at the second order in $\varepsilon$ are
 \begin{eqnarray}
   \frac{\partial n_2}{\partial t_0} +\frac{\partial (u_2)}{\partial x_0} ={\cal F}_1,
\nonumber \\
\label{eq_i10}
   \frac{\partial u_2}{\partial t_0} +\frac{\partial \phi_2}{\partial x_0} ={\cal F}_2,
\\
   \frac{\partial^2 \phi_2}{\partial x_0^2} -c_1 \phi_2 +n_2 ={\cal F}_3,
\nonumber
\end{eqnarray}
where ${\cal F}_1$, ${\cal F}_2$, and ${\cal F}_3$ are known functions of 
$\Psi_j$ which are given in Appendix B. To solve Eqs. (\ref{eq_i10}), 
we use the following {\em ansatz}
\begin{eqnarray}
\label{eq12.2}
   S_2 =S_{2}^{(0)} 
       +\left[ S_{2,1}^{(1)} e^{i \theta_1} +S_{2,1}^{(2)} e^{2 i \theta_1}
               +S_{2,2}^{(1)} e^{i \theta_2} +S_{2,2}^{(2)} e^{2 i \theta_2} \right.
\nonumber \\
\left.
               +S_{2,+}^{(1)} e^{i (\theta_1 +\theta_2)} 
               +S_{2,-}^{(1)} e^{2 i (\theta_1 -\theta_2)} +c.c.
        \right],
\end{eqnarray}
where $S =n, u$ or $\phi$; by ``$c.c$" we have denoted the complex conjugate of the 
expression within the curly brackets. Upon  substitution of Eqs. (\ref{eq12.2}) 
into Eqs. (\ref{eq_i10}), we obtain the systems of equations 
\begin{eqnarray}
   -\omega_j n_{2,j}^{(1)} +k_j u_{2,j}^{(1)} =\mu_{1,j}, 
\nonumber \\
   -\omega_j u_{2,j}^{(1)} +k_j \phi_{2,j}^{(1)} =\mu_{2,j}, 
\label{eq12.3}
\\
    -(k_j^2 +c_1) \phi_{2,j}^{(1)} +n_{2,j}^{(1)} =\mu_{3,j},
\nonumber
\end{eqnarray}
resulting from all terms proportional to $e^{i \theta_1}$ and $e^{i \theta_2}$, 
respectively (to be discussed later; note that $\mu_{1,j}$, $\mu_{2,j}$, and 
$\mu_{3,j}$, involving temporal and spatial derivatives of $\Psi_j$, for $j=1,2$, 
are given in Appendix B), 
\begin{eqnarray}
   -2 \omega_j n_{2,j}^{(2)} +2 k_j u_{2,j}^{(2)} =-2 k_j n_{1,j}^{(1)} u_{1,j}^{(1)}, 
\nonumber \\
\label{eq12.4}
   -2 \omega_j u_{2,j}^{(2)} +2 k_j \phi_{2,j}^{(2)} =-k_j u_{1,j}^{(1)}{^2},
\\
  -(4 k_j^2 +c_1) \phi_{2,j}^{(2)} +n_{2,j}^{(2)}=c_2 \Psi_j^2,
\nonumber
\end{eqnarray}
for $j=1$ and $2$, the latter resulting from the  terms proportional to 
$e^{2 i \theta_1}$ and $e^{2 i \theta_2}$, respectively. The second harmonic 
amplitudes are given by  solutions in the form 
\begin{eqnarray}
  n_{2,j}^{(2)} =C_{n,2,j}^{(2)} \Psi_j^2, 
\nonumber \\
\label{eq12.5}
  u_{2,j}^{(2)} =C_{u,2,j}^{(2)} \Psi_j^2,
\\
  \phi_{2,j}^{(2)} =C_{\phi,2,j}^{(2)} \Psi_j^2,
\nonumber
\end{eqnarray}
where the coefficients $C_{n,2,j}^{(2)}$, $C_{u,2,j}^{(2)}$, and 
$C_{\phi,2,j}^{(2)}$ are given in Appendix A, and finally
\begin{eqnarray}
\label{eq12.6}
   -(\omega_1 +\omega_2) n_{2,+}^{(1)} +(k_1 +k_2) u_{2,+}^{(1)} 
   =f_1^{(+)} \Psi_1 \Psi_2, 
   \\ 
   \label{eq12.7}
   -(\omega_1 +\omega_2) u_{2,+}^{(1)} +(k_1 +k_2) \phi_{2,+}^{(1)} 
   =f_2^{(+)} \Psi_1 \Psi_2,
   \\
   \label{eq12.8}
  -\left[(k_1+k_2)^2 +c_1 \right] \phi_{2,+}^{(2)} +n_{2,+}^{(1)}
  =f_3^{(+)} \Psi_1 \Psi_2,
\end{eqnarray}
and 
\begin{eqnarray}
\label{eq12.09}
   -(\omega_1 -\omega_2) n_{2,-}^{(1)} +(k_1 -k_2) u_{2,-}^{(1)} 
   =f_1^{(-)} \Psi_1 \Psi_2^\star, 
   \\ 
   \label{eq12.10}
   -(\omega_1 -\omega_2) u_{2,-}^{(1)} +(k_1 -k_2) \phi_{2,-}^{(1)} 
   =f_2^{(-)} \Psi_1 \Psi_2^\star,
   \\
   \label{eq12.11}
  -\left[(k_1 -k_2)^2 +c_1 \right] \phi_{2,-}^{(2)} +n_{2,-}^{(1)}=
  =f_3^{(-)} \Psi_1 \Psi_2^\star,
\end{eqnarray}
where $f_1^{(\pm)}$, $f_2^{(\pm)}$, and $f_3^{(\pm)}$ are functions of $k_1$ 
and $k_2$ and are given in Appendix B. 
The last two systems result from the terms proportional to 
$e^{i (\theta_1 +\theta_2)}$ and $e^{i (\theta_2 -\theta_2)}$, respectively. 
Their solution reads: 
\begin{eqnarray}
\label{eq12.13}
   n_{2,+}^{(1)} =C_{n,2,+}^{(1)} \Psi_1 \Psi_2, ~~~~ 
   n_{2,-}^{(1)} =C_{n,2,-}^{(1)} \Psi_1 \Psi_2^\star,
\\
\label{eq12.14}
   u_{2,+}^{(1)} =C_{u,2,+}^{(1)} \Psi_1 \Psi_2, ~~~~
   u_{2,-}^{(1)} =C_{u,2,-}^{(1)} \Psi_1 \Psi_2^\star,
\\
\label{eq12.14.2}
   \phi_{2,+}^{(1)} =C_{\phi,2,+}^{(1)} \Psi_1 \Psi_2, ~~~~
   \phi_{2,-}^{(1)} =C_{\phi,2,-}^{(1)} \Psi_1 \Psi_2^\star,
\end{eqnarray}
where again the coefficients $C_{n,2,\pm}^{(1)}$, $C_{u,2,\pm}^{(1)}$, and 
$C_{\phi,2,\pm}^{(1)}$ are given in Appendix A. 

In the same (second) order
($\sim \varepsilon^2$), we also obtain an equation involving the ``constant'' 
terms (zeroth order harmonics) in the form
\begin{equation}
\label{eq12.15}
   -c_1 \phi_2^{(0)} +n_2^{(0)} =+2 c_2 \left( |\Psi_1|^2 +|\Psi_2|^2 \right),
\end{equation} 
which will be used in the next subsection.

Before moving on to the equations which are third order in $\varepsilon$, we 
return to the discussion of the solution of Eqs. (\ref{eq12.3}). The latter, 
which actually consists of two $3\times 3$ uncoupled systems for the variables 
$\{ n_{2,1}^{(1)}, u_{2,1}^{(1)}, \phi_{2,1}^{(1)} \}$ and 
$\{ n_{2,2}^{(1)}, u_{2,2}^{(1)}, \phi_{2,2}^{(1)} \}$, can be written in 
matrix form as
\begin{eqnarray}
\label{eq130}
\begin{bmatrix}
   -\omega_j   & +k_j        &  0  \\
    0          & -\omega_j   & +k_j  \\
    1          & 0           & -(k_j^2 +c_1) 
\end{bmatrix}
\begin{bmatrix}
   n_{2,j}^{(1)}     \\
   u_{2,j}^{(1)}     \\
  \phi_{2,j}^{(1)}       
\end{bmatrix}
 =
\begin{bmatrix}
  \mu_{1,j}    \\
  \mu_{2,j}    \\
  \mu_{3,j}     
\end{bmatrix}
,
\end{eqnarray}
where $\mu_{m,j}$ ($m=1,2,3$ and $j=1,2$) are provided in Appendix B. Note that 
the determinants of the two inhomogeneous systems of Eqs. (\ref{eq13}) 
are zero, 
since this was the requirement for obtaining the frequency dispersion relations. 
Thus, in order to find nontrivial solutions for the two inhomogeneous systems 
of Eqs. (\ref{eq13}), it is imposed that the following determinants should also 
vanish: 
\begin{eqnarray}
\label{eq140}
D_j' =
\begin{vmatrix}
   \mu_{1,j}    & +k_j       &  0    \\
   \mu_{2,j}    & -\omega_1  & +k_j  \\
   \mu_{3,j}    & 0          & -(k_j^2 +c_1) 
\end{vmatrix}
=0.
\end{eqnarray}
These determinants result from the replacement of the first column of the 
determinant $D_j$ with the vector 
${\bf \mu}_j =[ \mu_{1,j} \, \mu_{2,j} \, \mu_{3,j} ]^T$. After calculations we 
obtain that in order for $D_j'$ to vanish, the following compatibility conditions 
\begin{equation}
\label{eq150}
   \frac{\partial \Psi_j}{\partial t_1} =
    -v_{g,j} \frac{\partial \Psi_j}{\partial x_1}
\end{equation} 
(for both $j = 1$ and $2$) should hold. The physical implication is that the (two) 
wavepacket envelopes travel at their respective group velocity. This was 
qualitatively expected from earlier works \cite{Kourakis2005,Infeld-Rowlands}.

\subsection{Derivation of a system of coupled NLS equations}

The equations in order $\varepsilon^3$ are
\begin{eqnarray}
   \frac{\partial n_3}{\partial t_0} +\frac{\partial (u_3)}{\partial x_0} 
   ={\cal G}_1,
\nonumber \\
\label{eq16}
   \frac{\partial u_3}{\partial t_0} +\frac{\partial \phi_3}{\partial x_0} 
   ={\cal G}_2,
\\
   \frac{\partial^2 \phi_3}{\partial x_0^2} -c_1 \phi_3 +n_3 
   ={\cal G}_3,
\nonumber
\end{eqnarray}
where ${\cal G}_1$, ${\cal G}_2$, and ${\cal G}_3$ are expressions involving lower ($\le 2$) 
order harmonics of $n$, $u$, and $\phi$, which  are given in Appendix B. 
We shall seek a multi-harmonic solution by substituting into the system of Eqs. (\ref{eq16}) 
the following {\em Ansatz}
\begin{eqnarray}
\label{eq20}
   S_3 = S_{3}^{(0)} +\left[] S_{3,1}^{(1)} e^{i \theta_1} 
        +S_{3,1}^{(2)} e^{2 i \theta_1} +S_{3,1}^{(3)} e^{3 i \theta_1}
        +S_{3,2}^{(1)} e^{i \theta_2} 
        +S_{3,2}^{(2)} e^{2 i \theta_2} +S_{3,2}^{(3)} e^{3 i \theta_2}
    \right.
\nonumber \\
\left.
    +S_{3,+}^{(1)} e^{i (\theta_1 +\theta_2)} 
    +S_{3,-}^{(1)} e^{2 i (\theta_1 -\theta_2)} 
    +S_{3,+12}^{(1)} e^{i (\theta_1 +2\theta_2)} 
    +S_{3,-12}^{(1)} e^{i (\theta_1 -2\theta_2)}
\right.
\nonumber \\
\left.
    +S_{3,+21}^{(1)} e^{i (2\theta_1 +\theta_2)} 
    +S_{3,-21}^{(1)} e^{i (2\theta_1 -\theta_2)} +c.c.
  \right],
\end{eqnarray}
where $S =n, u$, and $\phi$ and $c.c.$ denotes the complex conjugate of the 
expression within the curly brackets. 

Isolating the $\ell=0$ contributions, we obtain two 
equations for the zeroth harmonic amplitudes 
which can be combined 
with Eq. (\ref{eq12.15}) to provide a $3\times 3$ system to be satisfied by the (3) 
unknowns $n_2^{(0)}$, $u_2^{(0)}$, and $\phi_2^{(0)}$, i.e.
\begin{eqnarray}
\label{eq21}
   \frac{\partial n_2^{(0)}}{\partial t_1} 
   +\frac{\partial u_2^{(0)}}{\partial x_1} 
   &=&-2 \frac{\partial}{\partial x_1} 
    \left[ \left( \frac{k_1}{\omega_1} \right)^3 |\Psi_1|^2 
     +\left( \frac{k_2}{\omega_2} \right)^3 |\Psi_2|^2 
    \right],
\\
\label{eq22}
   \frac{\partial u_2^{(0)} }{\partial t_1} 
   +\frac{\partial \phi_2^{(0)}}{\partial x_1}
   &=&-\frac{\partial}{\partial x_1} 
   \left[ \left( \frac{k_1}{\omega_1} \right)^2 |\Psi_1|^2
    +\left( \frac{k_2}{\omega_2} \right)^2 |\Psi_2|^2 \right]
\\
   \label{eq23}
   n_2^{(0)} &=&c_1 \phi_2^{(0)} +2 c_2 
    \left( |\Psi_1|^2 +|\Psi_2|^2 \right) \, .
\end{eqnarray} In writing down the latter equations, we have also used 
the relations (\ref{eq_i09}). To  solve the $3\times 3$  system thus 
obtained for the zeroth harmonics, we shall introduce the {\em Ansatz} 
\begin{eqnarray}
\label{eq24.1}
  \phi_2^{(0)} =C_{\phi,2,1}^{(0)} |\Psi_1|^2 +C_{\phi,2,2}^{(0)} |\Psi_2|^2,
  \nonumber \\
\label{eq24.2}
  u_2^{(0)} =C_{u,2,1}^{(0)} |\Psi_1|^2 +C_{u,2,2}^{(0)} |\Psi_2|^2,
\\
  n_2^{(0)} =C_{n,2,1}^{(0)} |\Psi_1|^2 +C_{n,2,2}^{(0)} |\Psi_2|^2,
\nonumber
\end{eqnarray} 
into Eqs. (\ref{eq21})-(\ref{eq23}) and use Eq. (\ref{eq15}) to determine 
the coefficients $C_{\phi,2,j}$, $C_{u,2,j}$, and $C_{n,2,j}$ whose 
expressions are given in the Appendix A. 

Finally, the equations obtained (in third order in $\varepsilon$) for the 
(3rd-rder) first harmonic mode $\ell =1$ read
\begin{eqnarray}
  -i \omega_j n_{3,j}^{(1)} +i k_j u_{3,j}^{(1)} ={\cal R}_{1,j},
\nonumber \\
\label{eq25}
  -i \omega_j u_{3,j}^{(1)} +i k_j \phi_{3,j}^{(1)} ={\cal R}_{2,j},
\\
  -\left( k_j^2 +c_1 \right) \phi_{3,j}^{(1)} +n_{3,j}^{(1)} ={\cal R}_{3,j},
  \nonumber
\end{eqnarray}
where the functions ${\cal R}_{1,j}$, ${\cal R}_{2,j}$, and ${\cal R}_{3,j}$
are given in Appendix B.

A lengthy algebraic manipulation of Eqs. (\ref{eq25}) -- aimed at eliminating the 
variables $n_{3,j}^{(1)}$, $u_{3,j}^{(1)}$, and $\phi_{3,j}^{(1)}$ -- leads us to 
an equation in the form: 
\begin{equation}
\label{eq29}
  -i \omega_j {\cal R}_{1,j} -i k_j {\cal R}_{2,j} +\omega_j^2 {\cal R}_{2,j} =0,
\end{equation}
where the expressions for ${\cal R}_{m,j}$ ($m=1,2,3$ and $j=1,2$) are given by 
Eqs. (\ref{eq26})-(\ref{eq28}) in Appendix B.

Finally, by substituting the expressions for 
${\cal R}_{m,j}$ into Eq. (\ref{eq29}) and using the obtained solutions for the 
harmonic amplitudes 
$n_{1,j}^{(1)}$, $u_{1,j}^{(1)}$, $\phi_{1,j}^{(1)} =\Psi_j$, $n_{2,j}^{(2)}$, 
$u_{2,j}^{(2)}$, $\phi_{2,j}^{(2)}$, $n_{2}^{(0)}$, $u_{2}^{(0)}$, and 
$\phi_{2}^{(0)}$, we obtain after a tedious and lengthy calculation with  
equations
\begin{eqnarray}
\label{eq30}
   4 \omega_1 k_1 \frac{\partial^2 \Psi_1 }{\partial t_1 \partial x_1} 
  -\frac{k_1^2}{\omega_1^2} \frac{\partial^2 \Psi_1 }{\partial t_1^2}
  +(1-\omega_1^2) \frac{\partial^2 \Psi_1 }{\partial x_1^2}
  +2 i \frac{k_1^2}{\omega_1} \frac{\partial^2 \Psi_1 }{\partial t_2^2} 
  +2 i k_1 (1-\omega_1^2) \frac{\partial \Psi_1 }{\partial x_2}
  \nonumber \\
  +\left( \tilde{Q}_{11} |\Psi_1|^2 +\tilde{Q}_{12} |\Psi_2|^2 \right) \Psi_1 =0,
  \\
 \label{eq31}
   4 \omega_2 k_2 \frac{\partial^2 \Psi_2 }{\partial t_1 \partial x_1} 
  -\frac{k_2^2}{\omega_2^2} \frac{\partial^2 \Psi_2 }{\partial t_1^2}
  +(1-\omega_2^2) \frac{\partial^2 \Psi_2 }{\partial x_1^2}
  +2 i \frac{k_2^2}{\omega_2} \frac{\partial^2 \Psi_2 }{\partial t_2^2} 
  +2 i k_2 (1-\omega_2^2) \frac{\partial \Psi_2 }{\partial x_2}
  \nonumber \\
  +\left( \tilde{Q}_{21} |\Psi_1|^2 +\tilde{Q}_{22} |\Psi_2|^2 \right) \Psi_2 =0, 
\end{eqnarray}
where
\begin{equation}
\label{eq32}
   \tilde{Q}_{jj} =-2 \frac{k_j^3}{\omega_j} 
    \left( C_{u,2,j}^{(2)} +C_{u,2,j}^{(0)} \right)
   -k_j^2 \left( C_{n,2,j}^{(2)} +C_{n,2,j}^{(0)} \right)
   +2 c_2 \omega_j^2  
    \left( C_{\phi,2,j}^{(2)} +C_{\phi,2,j}^{(0)} \right) +3 c_3 \omega_j^2,
\end{equation}
and
\begin{eqnarray}
\label{eq33}
   \tilde{Q}_{12} =-2 \frac{k_1^3}{\omega_1} C_{u,2,2}^{(0)}     
             -k_1  \frac{k_2}{\omega_2} 
             \left( \omega_1 \frac{k_2}{\omega_2} +k_1 \right)
                    \left( C_{u,2,+}^{(1)} +C_{u,2,-}^{(1)} \right)
\nonumber \\
          -k_1^2 C_{n,2,2}^{(0)} -\omega_1 k_1 \frac{k_2}{\omega_2} 
                   \left( C_{n,2,+}^{(1)} +C_{n,2,-}^{(1)} \right)
\nonumber \\
    +2 c_2 \omega_1^2  \left( C_{\phi,2,2}^{(0)} +C_{\phi,2,+}^{(1)} 
    +C_{\phi,2,-}^{(1)} \right) 
    +6 c_3 \omega_1^2,
                        \\
\label{eq34}
   \tilde{Q}_{21} =-2 \frac{k_2^3}{\omega_2} C_{u,2,1}^{(0)}     
                   -k_2  \frac{k_1}{\omega_1} \left( \omega_2 \frac{k_1}{\omega_1} +k_2 \right)
\nonumber \\
                        \left( C_{u,2,+}^{(1)} +C_{u,2,-}^{(1)} \right)
                         -k_2^2 C_{n,2,1}^{(0)} -\omega_2 k_2 \frac{k_1}{\omega_1} 
                                  \left( C_{n,2,+}^{(1)} +C_{n,2,-}^{(1)} \right)
                         \nonumber \\
                   +2 c_2 \omega_2^2  \left( C_{\phi,2,1}^{(0)} +C_{\phi,2,+}^{(1)} 
                   +C_{\phi,2,-}^{(1)} \right) 
                   +6 c_3 \omega_2^2.  
\end{eqnarray}
Note that the terms proportional to the variables $n_{2,j}^{(1)}$, $u_{2,j}^{(1)}$, 
and $\phi_{2,j}^{(1)}$ were eliminated with the help of Eqs. (\ref{eq12.3}).

From Eqs. (\ref{eq30}) and (\ref{eq31}), it is straightforward to obtain a pair of coupled 
NLS equations in the familiar form 
\begin{eqnarray}
\label{eq35}
   i \left( \frac{\partial \Psi_1}{\partial t_2} 
    +v_{g,1} \frac{\partial \Psi_1}{\partial x_2} \right)
   +P_1 \frac{\partial^2 \Psi_1}{\partial x_1^2}
   +\left(Q_{11} |\Psi_1|^2 +Q_{12} |\Psi_2|^2 \right) \Psi_1 =0, 
\\
\label{eq36}
   i \left( \frac{\partial \Psi_2}{\partial t_2} 
    +v_{g,2} \frac{\partial \Psi_2}{\partial x_2} \right)
    +P_2 \frac{\partial^2 \Psi_2}{\partial x_1^2}
    +\left(Q_{21} |\Psi_1|^2 +Q_{22} |\Psi_2|^2 \right) \Psi_2 =0 \, , 
\end{eqnarray}
where (for $j=1,2$) 
\begin{equation}
\label{eq37}
   P_j =-\frac{3}{2} c_1 \frac{k_j}{ (k_j^2 +c_1)^{5/2} }, 
\end{equation} 
are the (2) dispersion coefficients (analogous to the group-velocity-dispersion/GVD 
terms in nonlinear optics; note that 
$P_j = \frac{1}{2}\frac{\partial \omega_j}{\partial k_j}$, and 
\begin{equation}
\label{eq38}
    Q_{jj} = \frac{\omega_j}{2 k_j^2} \tilde{Q}_{jj}, ~~~~
    Q_{12} = \frac{\omega_1}{2 k_1^2} \tilde{Q}_{12}, ~~~~
    Q_{21} = \frac{\omega_2}{2 k_2^2} \tilde{Q}_{21}
\end{equation} 
are the self-modulation (self-nonlinearity) coefficient and the (2) 
cross-modulation (coupling nonlinearity) coefficients, respectively. 
(The tilded quantities in the latter three expressions were defined above.)

\begin{figure}[!h]
    \centering
    \includegraphics[width=12cm]{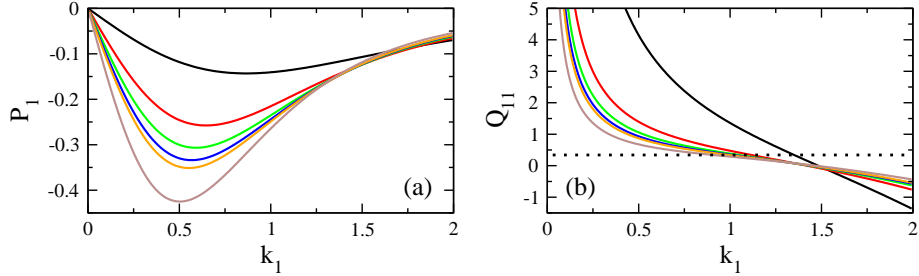}
    \caption{
    The frequency dispersion coefficient $P_1$, and the nonlinearity coefficient 
    $Q_{11}$, calculated from Eqs. (\ref{eq37}) and the first of Eqs. 
    (\ref{eq38}), as a function of the wavenumber $k_1$ for $\kappa=2$ (black), 
    $\kappa=3$ (red), $\kappa=4$ (green), $\kappa=5$ (blue), $\kappa=6$ (orange),
    $\kappa=100$ (brown).
    }
    \label{fig3}
\end{figure}
In Fig. \ref{fig3}, the dispersion coefficient $P_1$ and the nonlinearity 
coefficient $Q_{11}$ are shown as a function of the wavenumber of the first 
carrier wave $k_1$. Note that the corresponding quantities for the second carrier 
wave (i.e., $P_2$, $Q_{22}$, etc.) as a function of the wavenumber $k_2$ have 
exactly the same form. As evident in Fig. \ref{fig3}(a), for the model 
considered here, the dispersion coefficients $P_j$ ($j=1,2$) are negative for 
any $k_j$, converging to zero for large values of $k_j$. Also, for the interval 
of $k_1$ values shown in the figure, the magnitude of $P_1$ increases with 
increasing spectral parameter $\kappa$. In the Maxwell-Boltzmann case (i.e., 
corresponding practically to the value of $\kappa =100$), the minimum value 
of $P_1$ is located at $k_1 \simeq 0.5$, while it shifts to high values of 
$k_1$ with increasing $\kappa$. 

On the other hand, the nonlinearity coefficients $Q_{jj}$ can be either positive 
or negative, depending on the value of $k_j$, as can be seen in Fig. \ref{fig3}(b). 
In the Maxwell-Boltzmann case, that corresponds to very high values of the 
spectral index $\kappa$ of the kappa distribution of the electrons, the 
coefficients $Q_{jj}$ change sign at a particular value of the wavenumber of the 
carrier wave $k_j$, i.e., at $k_j \equiv k_r =1.47$, in agreement with earlier 
results \cite{Shimizu1972,Kakutani1974,Lazarides2023}.

\paragraph{Critical carrier wavenumbers for non-Maxwellian plasmas.}

Generally, for arbitrary $\kappa$, the symbol $k_r$ denotes the value of $k_1$ 
at which $Q_{jj}$ has a single root, i.e., $k_r$ is defined as that $k_1$ for 
which $Q_{jj} (k_j; \kappa) =0$. As it can be implied from the definition of 
$Q_{jj}$ (for either $j=1$ or 2), $k_r$ is a(n undisclosed) function of $\kappa$. 
\begin{figure}[!t]
    \centering
    \includegraphics[width=10cm]{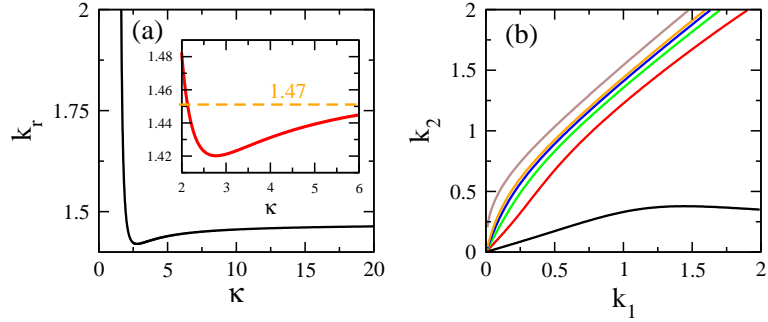}
    \caption{
    (a) The root $k_r$ of $Q_{11}$ as a function of the spectral index $\kappa$. 
    For large values of $\kappa$, that root converges asymptotically 
    to the value $k_1 =k_r =1.47$, i.e., its value in the Maxwell-Boltzmann 
    case.
    The inset shows a magnified part of the main figure for physically relevant
    values of $\kappa$. The orange-dashed line is located at $k_r =1.47$.
    (b) The zeroes of the coupling coefficient $Q_{12}$ on the $k_1 - k_2$ plane are shown 
    (in ascending order, top to bottom)
    for $\kappa=2$ (black), $\kappa=3$ (red), $\kappa=4$ (green), $\kappa=5$ 
    (blue), $\kappa=6$ (orange), $\kappa=100$ (brown). 
    The coupling coefficient $Q_{12}$ is negative in the area of the 
    $k_1 - k_2$ plane above the corresponding curve (and positive underneath). 
    Note that the coupling coefficient $Q_{21}$ can be obtained from $Q_{12}$ by 
    interchanging $k_1$ and $k_2$ in the corresponding expressions. The corresponding 
    curves for the zeros of $Q_{21}$ can be obtained by a reflection with respect to 
    the diagonal $k_1 =k_2$ (and thus $Q_{21}$ would be positive in the area above 
    the curves, and negative underneath). 
    }
    \label{fig4}
\end{figure}

The dependence of the root $k_r$ on the spectral index $\kappa$ is shown in Fig.
\ref{fig4}(a) for a large interval of $\kappa$ values. As it can be observed in 
Fig. \ref{fig4}(a), the root $k_r$ diverges as $\kappa \rightarrow 3/2$. 
The value of the root $k_r$ decreases rapidly as $\kappa$ increases from the 
value $3/2$, then reaches a minimum around $\kappa \simeq 2.5$ close to the 
border of far- to near-equilibrium, and then, with further increasing $\kappa$, 
it increases slowly to converge to $k_r =1.47$ from below for very large values 
of $\kappa$ (Maxwell-Boltzmann regime).

The coupling coefficients $Q_{12}$ and $Q_{21}$, on the other hand, depend on 
both $k_1$ and $k_2$ as can be observed from Eqs. (\ref{eq38}). It turns out 
that, on the $k_1 -k_2$ plane, these coefficients may have either positive or 
negative values. The boundary between the regions corresponding to opposite signs 
for $Q_{12}$ is shown for several values of $\kappa$ in Fig. \ref{fig4}(b), 
in which $Q_{12}$ is negative in the areas of the $k_1 -k_2$ plane above the 
boundary. It is obvious from the above Figure that smaller values of the 2nd 
wavenumber ($k_2$) that might be associated with negative $Q_{12}$ for Maxwellian 
electrons (i.e. large $\kappa$) now switch to a positive $Q_{12}$ in the case of 
suprathermal electrons, i.e. for smaller $\kappa$. The analogous curves (boundaries) 
for $Q_{21}$ are obtained upon reflection with respect to the diagonal $k_1 - k_2$, 
hence in the areas above the boundaries $Q_{21}$ is positive (and negative underneath).

\begin{figure}[!t]
    \centering
    \includegraphics[width=10cm]{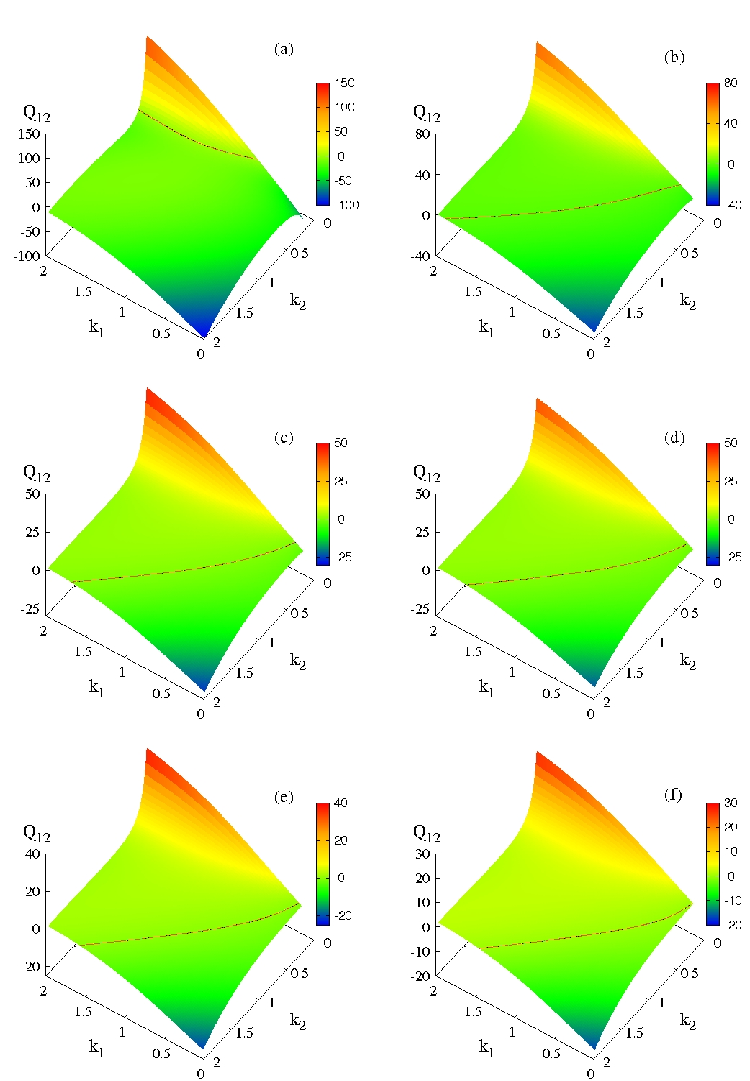}
    \caption{
    The nonlinear coupling coefficient $Q_{12}$ mapped on the plane of the 
    wavenumbers of the carrier waves $k_1$ and $k_2$, for 
    (a) $\kappa =2$,
    (b) $\kappa =3$,
    (c) $\kappa =4$,
    (d) $\kappa =5$,
    (e) $\kappa =6$,
    (f) $\kappa =100$. 
    In (f), the $\kappa$ distribution practically coincides with the 
    Maxwell-Boltzmann distribution.
    The black contour indicates the boundary between positive and 
    negative values of $Q_{12}$. Note that the plots above are shown from 
    $k_2 =0.1$ to $2$ when comparing the boundaries (the contours) with the 
    curves in Fig. \ref{fig4}(b). At low $k_2$ values, $Q_{12}$ becomes very 
    large and has been omitted, for clarity. The corresponding coupling 
    coefficients $Q_{21}$ are obtained upon a reflection around the line 
    $k_1 =k_2$ (i.e., upon a permutation between $k_1$ and $k_2$.).
    }
    \label{fig5}
\end{figure}

It is certainly useful to see not only these 
boundaries but all the values that $Q_{12}$ and $Q_{21}$ can take on the 
$k_1 -k_2$ plane. The first of them, i.e., the coefficient $Q_{12}$ is mapped 
on the $k_1 -k_2$ plane in Fig. \ref{fig5} for several values of $\kappa$. 
(Note that $Q_{12}$ and $Q_{21}$ exhibit reflection symmetry around $k_1 =k_2$.)
As it can be observed in Fig. \ref{fig5}, the coefficient $Q_{12}$ is mostly 
negative on the $k_1 -k_2$ plane for $\kappa =2$ (Fig. \ref{fig5}(a)), while 
the areas of negative and positive values are more or less of the same size for 
the other values of $\kappa$, i.e., for $\kappa =3,4,5,6,100$ 
(Figs. \ref{fig5}(b)-(f)). Also, it is clear that $Q_{12}$ takes very large 
positive or negative values in certain areas of the $k_1 -k_2$ plane. 
In particular, $Q_{12}$ is very large and positive for low $k_2$, while it is 
has very large negative values in the upper left corner of the $k_1 -k_2$ plane.
Actually, the plots of $Q_{12}$ are shown only for $k_2 > 0.1$ to avoid very 
large and divergent values. These large values of the coupling coefficients
$Q_{12}$ and $Q_{21}$ are probably responsible for the wide occurence of 
modulational instability in the CNLS system that will be discussed in the next 
section. Interestingly, only one (i.e. not both) of the $Q_{12}$ and $Q_{21}$ 
coefficients can become zero for a (any) particular pair of $k_1$ and $k_2$ 
values. In that case, for, say, $Q_{21} =0$, the second of the CNLS equations 
Eq. (\ref{eq36}) decouples from the first one, Eq. (\ref{eq35}), although the 
latter is still affected by the former \cite{Lazarides2023}.

\subsection{Reduced form of the CNLS system}

Equations (\ref{eq22}) and  (\ref{eq23}) can be transformed to a more common 
form of CNLS equations with a change of the independent variables, followed by 
a transformation of the wavefunctions $\Psi_1$ and $\Psi_2$. Let us introduce 
the new independent variables $\xi$ and $\tau$ through $\xi = x -v t$ and 
$\tau =t$, where the subscript in $x$ and $t$ have been dropped, 
and $v =( v_{g,1} +v_{g,2} )/2$ is the half-sum of the group velocities 
$v_{g,1}$ and $v_{g,2}$. By applying this change of variables to Eqs.  
(\ref{eq22}) and  (\ref{eq23}) we get 
\begin{eqnarray}
\label{eq_t05}
   i \left( \frac{\partial \Psi_1}{\partial \tau} 
   +\delta \frac{\partial \Psi_1}{\partial \xi} \right)
   +P_1 \frac{\partial^2 \Psi_1}{\partial \xi^2}
   +\left\{Q_{11} |\Psi_1|^2 +Q_{12} |\Psi_2|^2 \right\} \Psi_1 =0, 
\\
\label{eq_t06}
   i \left( \frac{\partial \Psi_2}{\partial \tau} 
   -\delta \frac{\partial \Psi_2}{\partial \xi} \right)
   +P_2 \frac{\partial^2 \Psi_2}{\partial \xi^2}
   +\left\{Q_{21} |\Psi_1|^2 +Q_{22} |\Psi_2|^2 \right\} \Psi_2 =0,
\end{eqnarray}
where $\delta =( v_{g,1} -v_{g,2} )/2$ is the half-difference of the group 
velocities. In writing the Eqs. (\ref{eq_t05}) and  (\ref{eq_t06}), it is 
implicitly understood that the differentiations involved in the dispersive 
terms are with respect to the variable $\xi_1$, while the higher-order 
differentiation in the beginning of the left-hand-side are with respect to 
$\xi_2$. The subscripts are nonetheless omitted, for simplicity, as they 
will not affect the final result.
Then, by applying to Eqs.  (\ref{eq_t05}) and  (\ref{eq_t06}) the 
transformation
$\Psi_j = \bar{\Psi}_j \exp\left[ i \left( \frac{\delta^2}{4 P_j} \tau 
-\frac{\delta}{2 P_j} \xi \right) \right]$, where $j=1,2$, we obtain
\begin{eqnarray}
\label{eq_t09}
   i \frac{\partial \bar{\Psi}_1}{\partial \tau} 
   +P_1 \frac{\partial^2 \bar{\Psi}_1}{\partial \xi^2}
   +\left\{Q_{11} |\bar{\Psi}_1|^2 +Q_{12} |\bar{\Psi}_2|^2 \right\} \bar{\Psi}_1 =0, 
\\
\label{eq_t10}
   i \frac{\partial \bar{\Psi}_2}{\partial \tau} 
   +P_2 \frac{\partial^2 \bar{\Psi}_2}{\partial \xi^2}
   +\left\{Q_{21} |\bar{\Psi}_1|^2 +Q_{22} |\bar{\Psi}_2|^2 \right\} \bar{\Psi}_2 =0,
\end{eqnarray}
where $\bar{\Psi}_j$ are complex functions of the new variables $\xi$ and $\tau$.

It should be noted that CNLS equations analogous (and formally equivalent) to either 
Eqs. (\ref{eq35})-(\ref{eq36}) or to Eqs. (\ref{eq_t09}) and (\ref{eq_t10}) above, 
have been obtained in various physical contexts earlier, describing 
nonlinear processes including pulse propagation in water waves \cite{He2022}, 
deep water wave propagation \cite{Ablowitz2015}, 
soliton propagation in left-handed transmission lines \cite{Veldes2013},
pulse propagation in optical nonlinear media \cite{Kivshar1993},
emergence of vector solitons in left-handed metamaterials \cite{Lazarides2005},
propagation of pulses with different polarization in anisotropic dispersive 
media \cite{Tyutin2022}. Analogous systems of equations have been obtained from 
various plasma fluid models \cite{Spatschek1978,Som1979,
McKinstrie1989,McKinstrie1990,Luther1990,Luther1992,Singh2013,Borhanian2017,
Lazarides2023}, but never in the case of a (non-symmetric) pair of electrostatic 
wavepackets (i.e. with different wavenumbers and amplitude), which is the focus 
of this study.

\section{Modulational instability analysis - analytical setting \label{section3}}

Modulational instability (MI) analysis for two co-propagating pulses in the 
plasma fluid described by Eqs. (\ref{eq01}) can be performed on the CNLS equations 
(\ref{eq_t05}) and (\ref{eq_t06}) following the procedure in Refs.
\cite{Borhanian2017,Kourakis2006}. We shall first seek a 
plane-wave solution for the system, which turns out to be of the form 
\begin{equation}
\label{mi01}
   \Psi_j =\Psi_{j,0} \, e^{i \tilde{\omega}_j \tau},
\end{equation}
where $\Psi_{j,0}$ ($j=1,2$) is  a constant amplitude 
(here assumed to be real) and $\tilde{\omega}_j$ 
is a frequency, to be determined. By inserting Eq. (\ref{mi01}) 
into Eqs. (\ref{eq_t05}) and (\ref{eq_t06}), we obtain
\begin{eqnarray}
\label{mi02}
   \tilde{\omega}_1 =Q_{11} \Psi_{1,0}^2 +Q_{12} \Psi_{2,0}^2, 
   \\
\label{mi022}   
   \tilde{\omega}_2 =Q_{21} \Psi_{1,0}^2 +Q_{22} \Psi_{2,0}^2.
\end{eqnarray}
Equations (\ref{mi01}) with the amplitude-dependent frequencies given in 
Eqs. (\ref{mi02}) and (\ref{mi022}) provide a nonlinear mode of the CNLS system 
wherein the propagation of waves is governed by Eqs. (\ref{eq_t05}) and 
(\ref{eq_t06}). The stability behavior of these modes against a small perturbation
can be addressed through standard MI analysis. Let us perturb the nonlinear 
modes by adding a small complex quantity $\varepsilon_j$ to their amplitudes 
($|\varepsilon_j| \ll |\Psi_{j,0}|$), as
\begin{equation}
\label{mi03}
  \Psi_j =( \Psi_{j,0} +\varepsilon_j ) \, e^{i \tilde{\omega}_j \tau}.
\end{equation}
By inserting the perturbed $\Psi_j$ into Eqs. (\ref{eq_t05}) and 
(\ref{eq_t06}) we obtain 
\begin{eqnarray}
\label{mi04}
 i \left(\frac{\partial \varepsilon_1}{\partial \tau} 
 +\delta \frac{\partial \varepsilon_1}{\partial \xi} \right)
 +P_1 \frac{\partial^2 \varepsilon_1}{\partial \xi^2} 
 +Q_{11} \Psi_{1,0}^2 (\varepsilon_1 +\varepsilon_1^\star) 
 +Q_{12} {\Psi}_{1,0} \Psi_{2,0} (\varepsilon_2 +\varepsilon_2^\star) =0, \\
 \label{mi05}
 i \left(\frac{\partial \varepsilon_2}{\partial \tau} 
 -\delta \frac{\partial \varepsilon_2}{\partial \xi} \right)
  +P_2 \frac{\partial^2 \varepsilon_2}{\partial \xi^2} 
 +Q_{22} \Psi_{2,0}^2 (\varepsilon_2 +\varepsilon_2^\star) 
 +Q_{21} \Psi_{2,0} \Psi_{1,0} (\varepsilon_1 +\varepsilon_1^\star) =0.
 \end{eqnarray}  
Then, by setting $\varepsilon_j =g_j +i h_j$, where $g_j$ and $h_j$ are real 
functions, and subsequently substituting into Eqs. (\ref{mi04}) and (\ref{mi05}) 
and eliminating $h_j$ from the equations, we get
\begin{eqnarray}
\label{mi06}
\left\{ \left(\frac{\partial}{\partial \tau} 
   +\delta \frac{\partial}{\partial \xi} \right)^2 
   +P_1 \left( P_1 \frac{\partial^2}{\partial \xi^2} 
   +2 Q_{11} \Psi_{1,0}^2 \right) \frac{\partial^2}{\partial \xi^2} \right\} g_1
\nonumber \\
   +2 P_1 Q_{12} \Psi_{2,0} \Psi_{1,0}  \frac{\partial^2}{\partial \xi^2} g_2 =0 \, , 
 \\
 \label{mi07}
 \left\{ \left(\frac{\partial}{\partial \tau} 
 -\delta \frac{\partial}{\partial \xi} \right)^2 
   +P_2 \left( P_2 \frac{\partial^2}{\partial \xi^2} 
   +2 Q_{22} \Psi_{2,0}^2 \right) \frac{\partial^2}{\partial \xi^2} \right\} g_2
\nonumber \\
   +2 P_2 Q_{21} \Psi_{1,0} \Psi_{2,0}  \frac{\partial^2}{\partial \xi^2} g_1 =0 \, .
 \end{eqnarray}

Eventually, by setting 
\begin{equation}
\label{mi08}
   g_j =g_{j,0} \, e^{i (K \xi -\Omega \tau)} +c.c ,
\end{equation}
where $c.c.$ denotes the complex conjugate,
and then substituting into the earlier equations, we obtain after 
some algebra the {\em compatibility condition} 
\begin{equation}
\label{mi09}
   \left[ (\Omega -\delta K)^2 -\Omega_1^2 \right] 
   \left[ (\Omega +\delta K)^2 -\Omega_2^2 \right] =\Omega_c^4,
\end{equation}
where ($j=1,2$)
\begin{eqnarray}
\label{mi10}
  \Omega_j^2 &=&P_j K^2 \left( P_j K^2 -2 Q_{jj} \Psi_{j,0}^2 \right),
\\ 
\label{mi10.2}
  \Omega_c^4 &=&4 P_1 P_2 Q_{12} Q_{21} \Psi_{1,0}^2 \Psi_{2,0}^2 K^4 .      
\end{eqnarray}

\begin{figure}[!t]
    \centering
    \includegraphics[width=10cm]{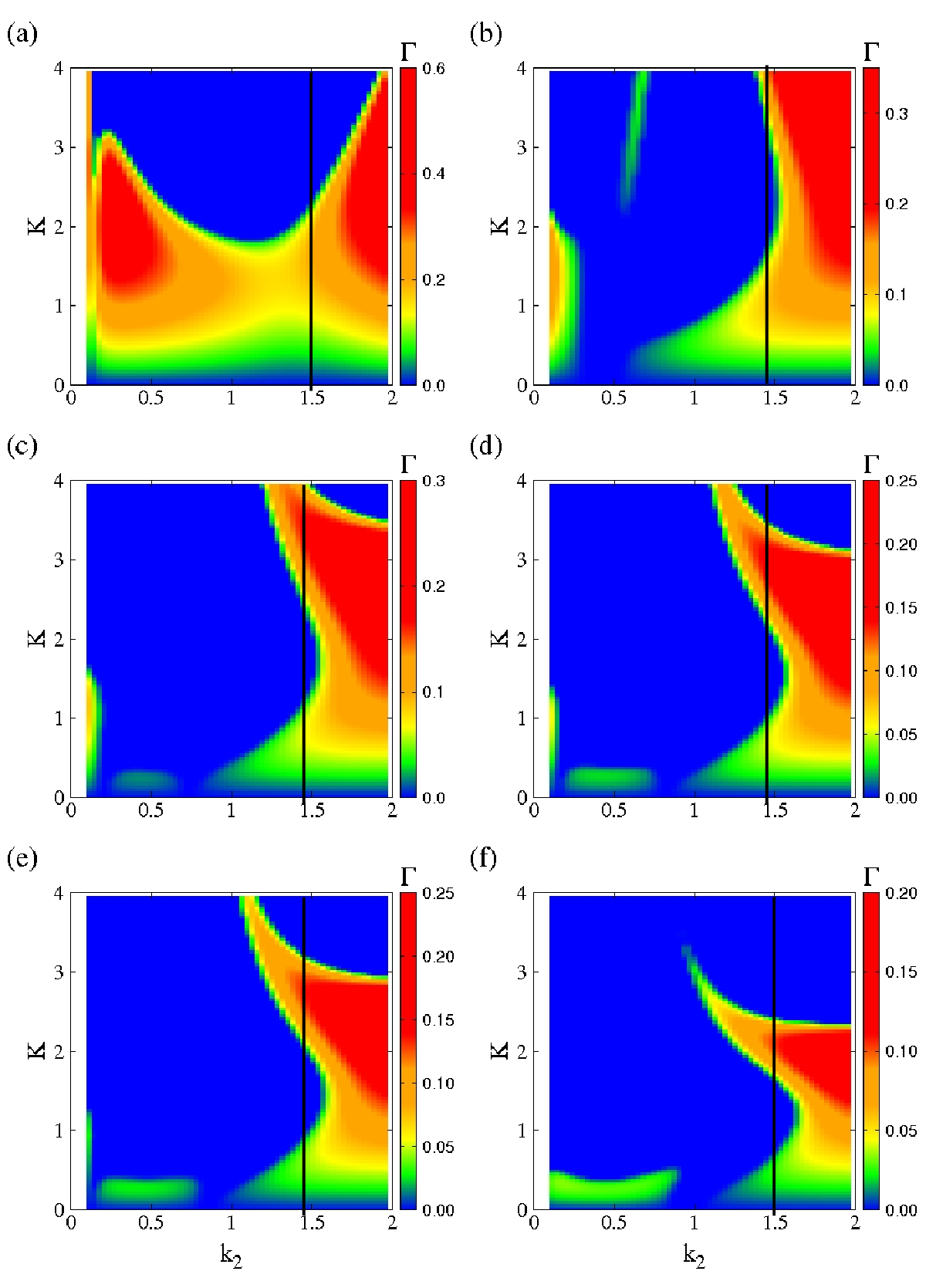}
    \caption{
    Maps of the growth rate $\Gamma$ on the plane of the wavenumber 
    of the second carrier wave $k_2$ and the wavenumber of the perturbation 
    $K$ for $k_1 =0.4$, $\Psi_{1,0} =\Psi_{2,0} =0.15$, and 
    (a) $\kappa =2$;
    (b) $\kappa =3$;
    (c) $\kappa =4$;
    (d) $\kappa =5$;
    (e) $\kappa =6$;
    (f) $\kappa =100$.
    For the selected value of $k_1$, the two single NLS equations resulting from
    decoupling the CNLS equations (e.g., by setting $Q_{12} =Q_{21} =0$ in Eqs. 
    (\ref{eq_t05}) and (\ref{eq_t06}) are either both in the stable regime (for
    $k_2 < k_r$ where $P_1 Q_{11} < 0$ and $P_2 Q_{22} < 0$), or the first is in
    the stable regime and the second in the unstable (for $k_2 > k_r$ where 
    $P_1 Q_{11} < 0$ and $P_2 Q_{22} > 0$). Note that $k_r$ depends on the 
    particular value of the spectral index $\kappa$ in accordance with the
    inset of Fig. \ref{fig4}(a).
    }
    \label{fig6}
\end{figure}
For modulational instability to set in, Eq. (\ref{mi09}) must have at least
one pair of complex conjugate roots, so that the (positive) imaginary part 
of these roots can be identified as the {\em growth rate} of the modulationally
unstable modes $\Gamma$. When Eq. (\ref{mi09}) has two pairs of complex 
conjugate roots, then the largest of their imaginary parts is identified as the 
growth rate of the modulationally unstable modes, $\Gamma$, i.e.,
\begin{equation}
\label{mi11}
   \Gamma =max[Im(\Omega_r)],       
\end{equation}
where $\Omega_r$ denotes one (each) of the four roots of the compatibility condition.
\begin{figure}[!t]
    \centering
    \includegraphics[width=10cm]{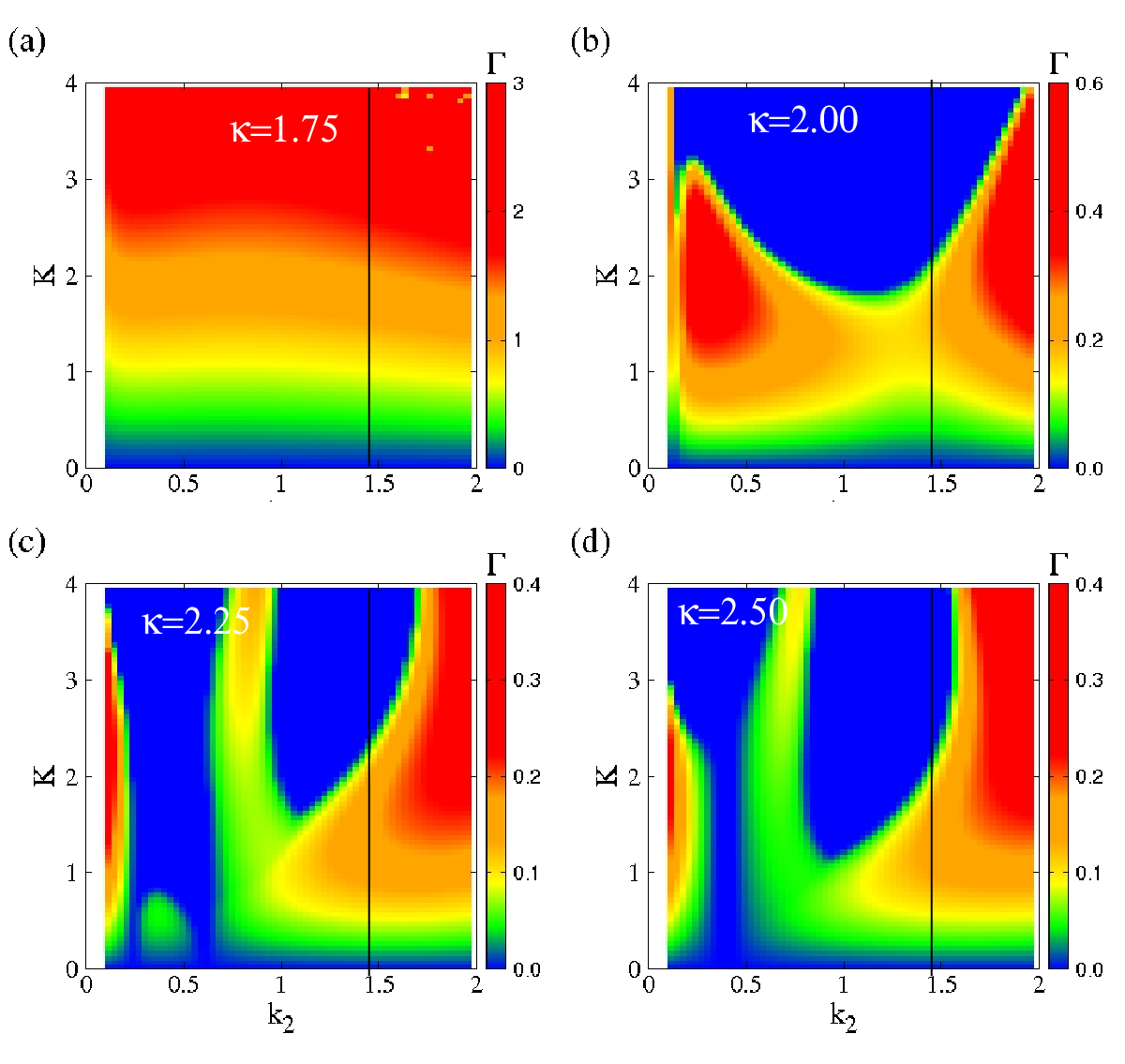}
    \caption{
    Maps of the growth rate $\Gamma$ on the $k_2 - K$ plane for (a) $\kappa =1.75$; 
    (b) $\kappa =2.00$; (c) $\kappa =2.25$; (d) $\kappa =2.50$.
    The other parameters as in Fig. \ref{fig6}. The vertical black-solid line
    indicates the boundary between two different regimes of the decoupled CNLS
    equations, i.e., the stable-stable regime for which $P_1 Q_{11} < 0$ and 
    $P_2 Q_{22} < 0$ ($k_2 < k_r$, left from the vertical line) and the 
    stable-unstable regime for which $P_1 Q_{11} < 0$ and $P_2 Q_{22} > 0$
    ($k_2 > k_r$, right from the vertical line). Note that $k_r$ depends weakly
    on the spectral index $\kappa$, but the differences cannot be observed in
    this scale.
    }
    \label{fig6.2}
\end{figure}

Note that a similar analysis may be applied to incoherent pairs of CNLS equations 
\cite{Huang2022} as well as to higher dimensional, non-autonomous CNLS equations 
\cite{Patel2021}.

\paragraph{Decoupled envelopes.}   
In order to facilitate later discussion, we consider now the special case in 
which the CNLS equations (\ref{mi06}) and (\ref{mi07}) are decoupled, i.e., 
the case in which either $Q_{12}$ or $Q_{21}$ (or even both) are equal to zero. In
that case, the compatibility conditions Eq. (\ref{mi09}) simplifies significantly,
and the perturbation frequencies $\Omega$ as a function of the perturbation 
wavenumber $K$ are given by 
\begin{equation}
\label{mi12}
    \Omega =+\delta K \pm \Omega_1, ~~~~\Omega =-\delta K \pm \Omega_2,
\end{equation}
for the first and the second NLS equation that results from decoupling Eqs.
(\ref{mi06}) and (\ref{mi07}), respectively. Correspondingly, the criterion for
modulational instability in the first and second decoupled equation becomes,
respectively, $P_1 Q_{11} < 0$ and $P_2 Q_{22} < 0$. Note that these products 
depend on $k_1$ only and $k_2$ only, respectively. Since $P_j$ ($j=1,2$) in the 
considered model is always negative, the sign of the product $P_j Q_{jj}$ 
depends here on $Q_{jj}$. Now, if we recall that $Q_{jj} > 0$ for 
$k_j < k_r (\kappa)$ while $Q_{jj} < 0$ for $k_j > k_r (\kappa)$, we arrive at
the following result: Within the considered model, the first 
NLS equation (for the 1st wavepacket) is modulationally stable for 
$k_1 < k_r (\kappa)$,  while it is modulationally unstable otherwise. An identical 
conclusion is drawn and can be formulated for the second wavepacket (if decoupled 
from the first), upon formally shifting $1 \rightarrow 2$, i.e. it will be 
stable for $k_2 < k_r (\kappa)$ and unstable otherwise. 

The conclusions in the latter paragraph recover what has long been known for the 
modulational stability profile of the NLS equation \cite{Sulem1999},  
summarized in that stability is prescribed for $P Q < 0$ and instability occurs 
otherwise.


\section{Modulational instability: numerical parametric analysis \label{section4}}

We have calculated the growth rate $\Gamma$ numerically for the CNLS equations by 
finding the roots of the polynomial in $\Omega$ function resulting from Eq. 
(\ref{mi09}), and subsequently identifying the one with the highest imaginary 
part. That highest imaginary part is plotted as a function of the wavenumber 
of the perturbation $K$ and the wavenumber of the second carrier wave $k_2$ 
for six values of the spectra index $\kappa$, i.e., for $\kappa =2,3,4,5,6,100$,
and fixed value of $k_1$ and the amplitudes of the nonlinear wave modes
$A_{1,0}$ and $A_{2,0}$ (the latter are fixed throughout the paper to 
$A_{1,0} =A_{2,0} =0.15$).

In Fig. \ref{fig6}, in particular, the growth rate $\Gamma$ is mapped on the 
$k_2 - K$ plane for $k_1 =0.4 < k_r (\kappa)$ and the six values of the spectral 
index $\kappa$ mentioned earlier. For that value of $k_1$, the first of the 
decoupled CNLS equations is modulationally stable ($P_1 Q_{11} < 0$), while the 
second one is also modulationally stable for $k_2 < k_r (\kappa)$ 
($P_2 Q_{22} < 0$) while it is modulationally unstable for $k_2 > k_r (\kappa)$ 
($P_2 Q_{22} > 0$). The instability boundary for the second decoupled CNLS 
equation is indicated in Fig. \ref{fig6} (as well as in Figs. \ref{fig7} and 
\ref{fig8} below) by the black-solid vertical line.
All the sub-figures Fig. \ref{fig6} reveal complex instability patterns, in 
which modulational instability appears both in areas of the $k_2 - K$ plane 
where $P_1 Q_{11} < 0$ and $P_2 Q_{22} < 0$ (stable-stable regime), as well as 
in areas where $P_1 Q_{11} < 0$ and $P_2 Q_{22} > 0$ (stable-unstable regime).
The growth rate $\Gamma$ has not been mapped for $k_2 < 0.1$ on the $k_2 - K$
plane in all sub-figures for clarity, since it acquires very large and even
divergent values in this areas.
\begin{figure}[!t]
    \centering
    \includegraphics[width=10cm]{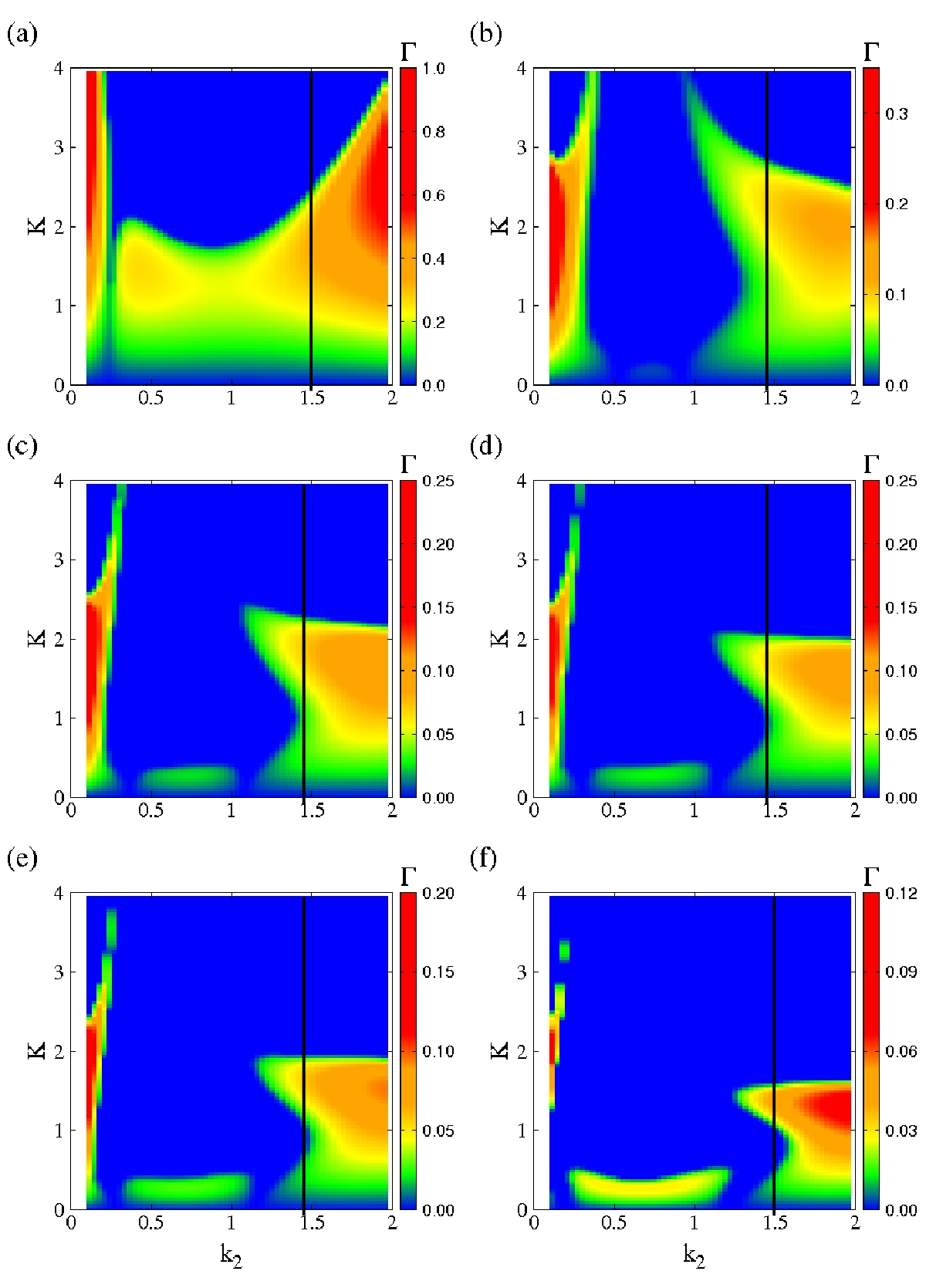}
    \caption{
    Maps of the growth rate $\Gamma$ on the plane of the wavenumber of the 
    second carrier wave $k_2$ and the wavenumber of the perturbation 
    $K$ for $k_1 =0.7$, $\Psi_{1,0} =\Psi_{2,0} =0.15$, and 
    (a) $\kappa =2$;
    (b) $\kappa =3$;
    (c) $\kappa =4$;
    (d) $\kappa =5$;
    (e) $\kappa =6$;
    (f) $\kappa =100$.
    For the selected value of $k_1$, the two single NLS equations resulting from
    decoupling the CNLS equations (e.g., by setting $Q_{12} =Q_{21} =0$ in Eqs. 
    (\ref{eq_t05}) and (\ref{eq_t06}) are either both in the stable regime (for
    $k_2 < k_r$ where $P_1 Q_{11} < 0$ and $P_2 Q_{22} < 0$), or the first is in
    the stable regime and the second in the unstable (for $k_2 > k_r$ where 
    $P_1 Q_{11} < 0$ and $P_2 Q_{22} > 0$). Note that $k_r$ depends on the 
    particular value of the spectral index $\kappa$ in accordance with the
    inset of Fig. \ref{fig4}(a).
    }
    \label{fig7}
\end{figure}

\begin{figure}[!t]
    \centering
    \includegraphics[width=10cm]{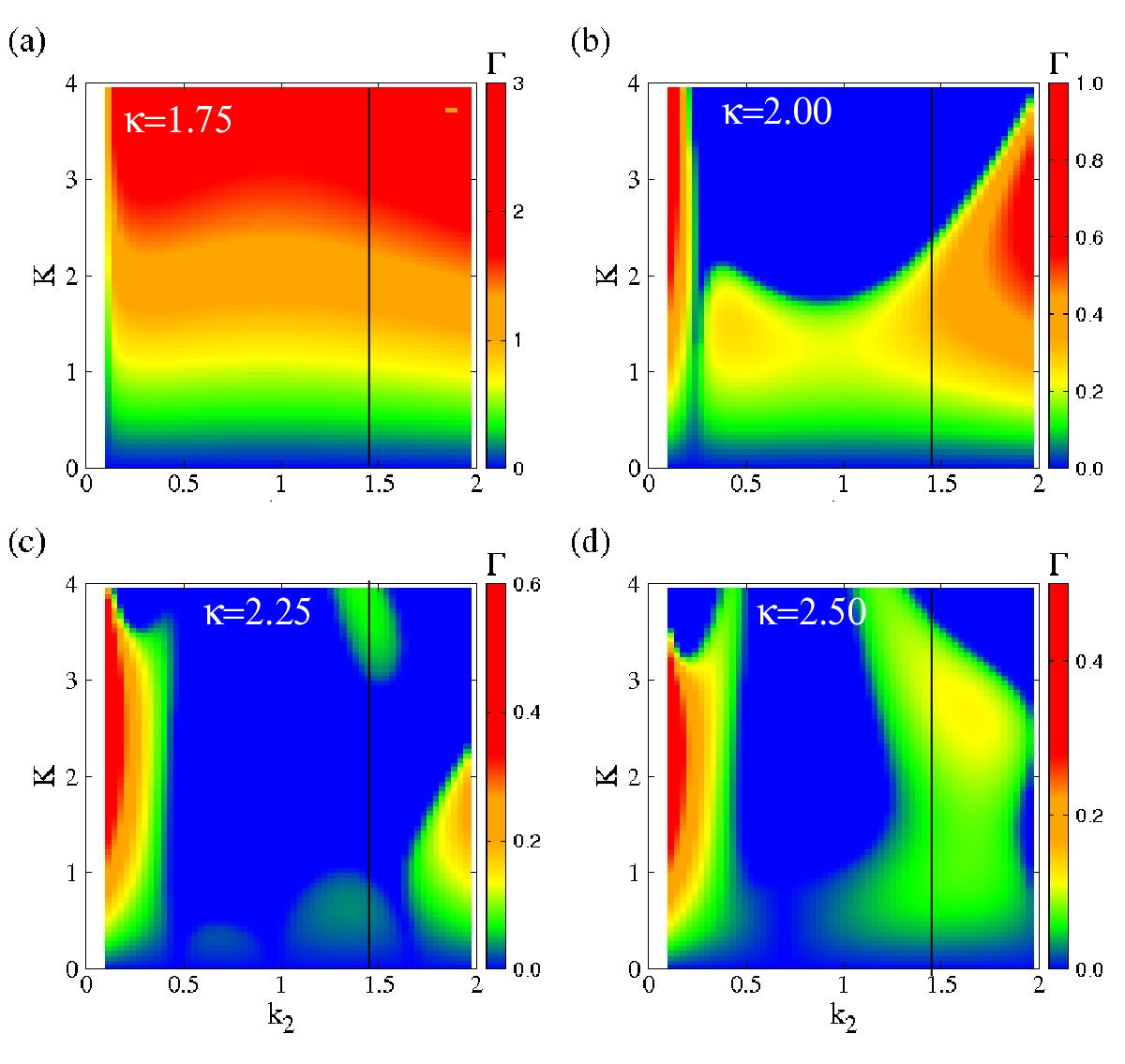}
    \caption{
    Maps of the growth rate $\Gamma$ on the $k_2 - K$ plane for (a) $\kappa =1.75$; 
    (b) $\kappa =2.00$; (c) $\kappa =2.25$; (d) $\kappa =2.50$.
    The other parameters as in Fig. \ref{fig7}. The vertical black-solid line
    indicates the boundary between two different regimes of the decoupled CNLS
    equations, i.e., the stable-stable regime for which $P_1 Q_{11} < 0$ and 
    $P_2 Q_{22} < 0$ ($k_2 < k_r$, left from the vertical line) and the 
    stable-unstable regime for which $P_1 Q_{11} < 0$ and $P_2 Q_{22} > 0$
    ($k_2 > k_r$, right from the vertical line). The stability situation in 
    these maps is the same as that in figs. \ref{fig6} and \ref{fig6.2}.
    Note that $k_r$ depends weakly on the spectral index $\kappa$, but the 
    differences cannot be observed in this scale.
    }
    \label{fig7.2}
\end{figure}
Generally speaking, the decrease of the spectral index $\kappa$ increases the
areas on the $k_2 - K$ plane in which MI occurs. Moreover, the maximum values
of the growth rate $\Gamma$ seem to increase considerably with decreasing 
$\kappa$ (note the different scales in the colorbars). Note also the difference 
in the mudulational instability patterns for values of $\kappa$ in the far- and
near-equilibrium region, i.e., for $\kappa =2$ and $\kappa =3,4,5,6,100$, 
respectively, which are separated by a rough empirical boundary at around 
$\kappa \simeq 2.5$. Indeed, for $\kappa =2$, the growth rate $\Gamma$ exhibits
very large values, and the CNLS system is modulationally unstable in the larger 
part of the $k_2 - K$ plane in which the decoupled CNLS equations are in the 
stable-stable regime ($P_1 Q_{11} < 0$ and $P_2 Q_{22} < 0$). For the next 
higher value of $\kappa$, i.e., for $\kappa =3$, the CNLS system is 
modulationally stable in the most part of the area in which the decoupled CNLS
equations are in the stable-stable regime. With further increasing the value of
$\kappa$, the areas of the $k_2 - K$ plane in which the CNLS system is 
modulationally unstable gradually shrink slightly, while the higher values of
the growth index $\Gamma$ also gradually decrease.

In Fig. \ref{fig6}, as well as in Figs. \ref{fig7} and \ref{fig8}, it is evident 
that the most significant variation in the patterns of $\Gamma$ 
seems to occur for values of the spectral index $\kappa$ within the 
strongly nonthermal  region, i.e., for $1.5 < \kappa < 2.5$. To the contrary, the 
variation of the $\Gamma$ pattern for $\kappa =3$ and up to infinity is rather
slight, at least as long as the size of the modulational instability areas on 
the $k_2 - K$ plane and the magnitude of $\Gamma$ in these areas are concerned.
In Fig. \ref{fig6.2}, maps of $\Gamma$ on the $k_2 - K$ plane are shown for more
values of $\kappa$ within the strongly nonthermal  region, i.e., for $\kappa =1.75$,
$2.25$, and $2.50$. In the figure, the map of the already shown map for 
$\kappa =2$ is also shown along with the other three. Inspection of the four
subfigures of Fig. \ref{fig6.2} readily reveals the dramatic variations of the
modulational instability $\Gamma$ patterns with the variation of $\kappa$.
For $\kappa =1.75$ (Fig. \ref{fig6.2}(a)), in particular, modulational 
instability accompanied by very high values of $\Gamma$ is apparent over almost 
the whole of the $k_2 - K$ plane shown. As we shall see also below, this is true 
for the other case studies presented here. Empirically we find that this the 
situation for $\kappa$ up to $1.90$, where where some modulationally stable 
islands begin to appear on the $k_2 - K$ plane. When $\kappa$ is increased to 
$2$, as we have also discussed in the previous figure, a substantial 
modulationally stable appear on the $k_2 - K$ plane that extends to both the
stable-stable and stable-unstable regions of the decoupled CNLS equations.
At the same time, the mudulationally unstable areas, which remain large, are
localized to two basic lobes, the one in the stable-stable and the other in the
stable-unstable region of the decoupled CNLS equations. In the centers of these 
lobes, the highest values of $\Gamma$ occur. With further increasing $\kappa$,
the modulationally stable area increases against the modulationally unstable 
ones. Moreover, the strongly modulationally unstable areas gradually 
concentrates in the stable-unstable region of the decoupled CNLS equations while
another strongly modulationally unstable area remains for low $k_2$. Remnants 
of the latter can be even observed in Fig. \ref{fig6}(b) (for $\kappa =3$).
\begin{figure}[!t]
    \centering
    \includegraphics[width=10cm]{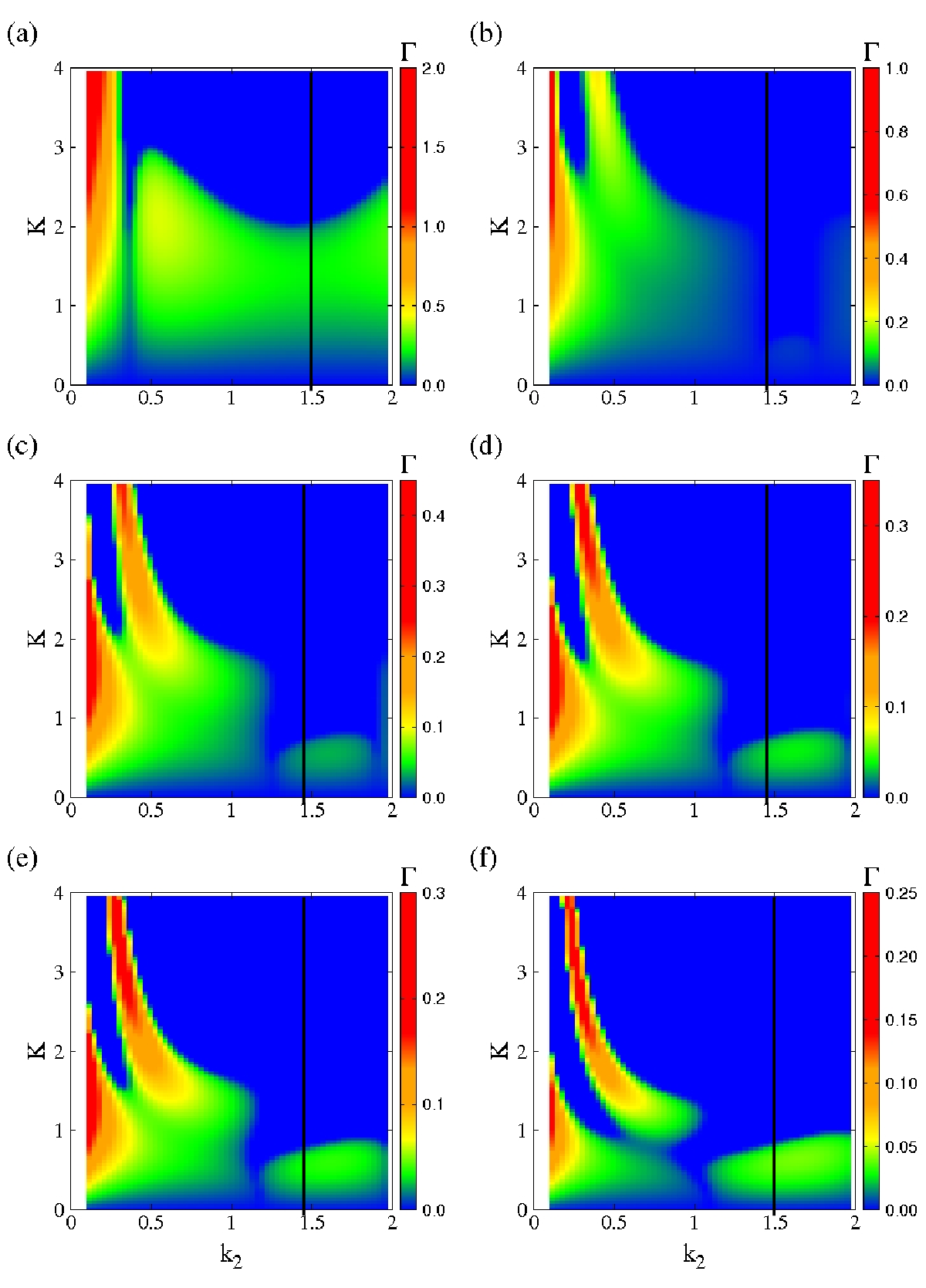}
    \caption{
    Maps of the growth rate $\Gamma$ on the plane of the wavenumber of the 
    second carrier wave $k_2$ and the wavenumber of the perturbation $K$ for 
    $k_1 =1.6$, $\Psi_{1,0} =\Psi_{2,0} =0.15$, and 
    (a) $\kappa =2$;
    (b) $\kappa =3$;
    (c) $\kappa =4$;
    (d) $\kappa =5$;
    (e) $\kappa =6$;
    (f) $\kappa =100$.
    For the selected value of $k_1$, the two single NLS equations resulting from
    decoupling the CNLS equations (e.g., by setting $Q_{12} =Q_{21} =0$ in Eqs. 
    (\ref{eq_t05}) and (\ref{eq_t06}) are either both in the unstable regime (for
    $k_2 > k_r$ where $P_1 Q_{11} > 0$ and $P_2 Q_{22} > 0$), or the first is in
    the unstable regime and the second in the stable (for $k_2 < k_r$ where 
    $P_1 Q_{11} > 0$ and $P_2 Q_{22} < 0$). Note that $k_r$ depends on the 
    particular value of the spectral index $\kappa$ in accordance with the
    inset of Fig. \ref{fig4}(a).
    }
    \label{fig8}
\end{figure}

The results shown in Fig. \ref{fig7} have been obtained with the same parameters 
as those used in Fig. \ref{fig6}, except that in this case $k_1 =0.7$. Thus, as
long as the stability of the decoupled CNLS equations is concerned, the situation
is exactly the same as that in Fig. \ref{fig6}, i.e., the decoupled CNLS equations
are in the stable-stable regime ($P_1 Q_{11} < 0$ and $P_2 Q_{22} < 0$) for
$k_2 < k_r (\kappa)$, while they are in the stable-unstable regime for 
$k_2 > k_r (\kappa)$ ($P_1 Q_{11} < 0$ and $P_2 Q_{22} > 0$). In comparison with
Fig. \ref{fig6}, we observe that for $\kappa =3,4,5,6,100$, areas of strong 
instability with relatively high values of the growth rate $\Gamma$ occur at
low $k_2$ for most of the perturbation wavenumbers shown in these figures. Again,
these strongly unstable areas shrink slightly with increasing spectral index from
$\kappa =3$ to $100$. A similar situation can be described for $\kappa =2$, where 
high values of the growth rate $\Gamma$ occur at low $k_2$ as in the case of 
higher $\kappa$; however, the values of $\Gamma$ are significantly higher in this
case (note again the difference in the scales of colorbars). Also, in a large 
part of the stable-stable area of the $k_2 - K$ plane modulational instability 
with moderate values of $\Gamma$ occurs. Generally speaking, the strength of the 
modulational instability is weaker and the areas where that instability occurs
are smaller in Fig. \ref{fig7} for $k_1 =0.7$ than in Fig. \ref{fig6} for 
$k_1 =0.4$ (except at low $k_2$). This can be observed more clearly for 
$\kappa \geq 3$.

As above, we also present maps of $\Gamma$ for a few more values of $\kappa$ in
the strongly nonthermal  region, i.e., in the interval $1.5 < \kappa < 2.5$ in Fig. 
\ref{fig7.2}. Roughly, the remarks made for the patterns of $\Gamma$ in Fig. 
\ref{fig6.2} also hold here. For $\kappa =1.75$ (Fig. \ref{fig7.2}(a)), almost 
all of the shown area of the $k_2 - K$ plane is modulationally unstable, with 
the growth rate achieving very high values of the same order as those in Fig. 
\ref{fig6.2}(a). Again, with increasing $\kappa$, the modulationally stable 
areas on the $k_2 - K$ plane grow to the expense of the modulationally unstable 
ones. The pattern of $\Gamma$ in Fig. \ref{fig7.2}(b) is similar to that in 
in Fig. \ref{fig6.2}(b) with the two modulationally unstable lobes and the 
strongly modulationally unstable area for low $k_2$. In Fig. \ref{fig7.2}(c),
the two lobes disappear, while a strongly modulationally unstable area at low 
$k_2$ remains, along with another modulationally unstable area which is located 
in the stable-unstable region of the decoupled CNLS equations, i.e., in the
area with $k_2 > k_r (\kappa)$. Peculiarly, the modulationally unstable areas 
in Fig. \ref{fig7.2}(d) are larger to those in Fig. \ref{fig7.2}(c). However,
the general tendency of shrinking of the modulationally unstable areas with
increasing $\kappa$ continuous for $\kappa > 2.5$. Note that in this case the 
strongly modulationally unstable area for low $k_2$ persist, although becoming 
much smaller with increasing $\kappa$, even for very large values of $\kappa$,
i.e., $\kappa =100$ (Fig. \ref{fig7}(f)).
\begin{figure}[!t]
    \centering
    \includegraphics[width=12cm]{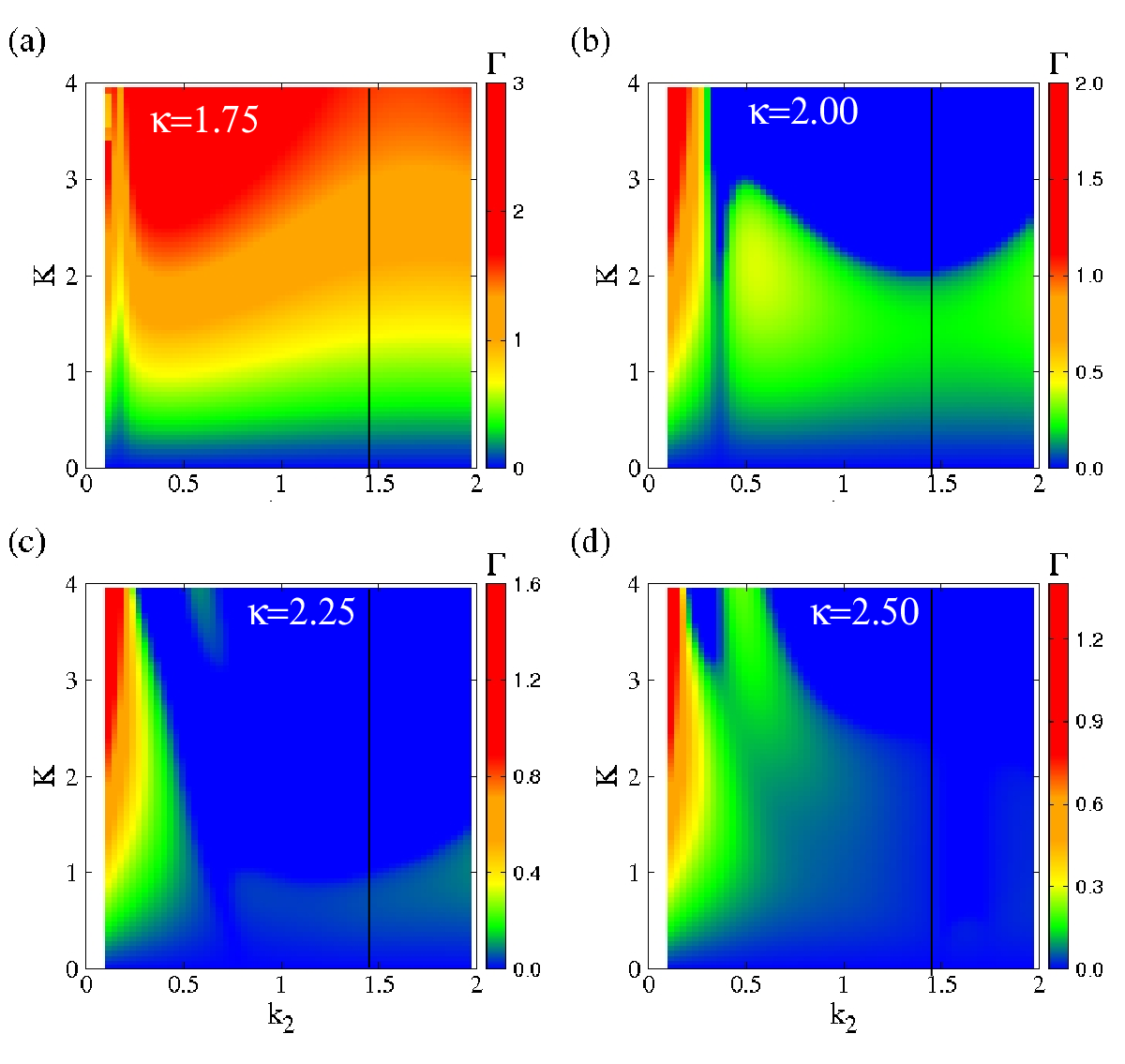}
    \caption{
    Maps of the growth rate $\Gamma$ on the $k_2 - K$ plane for (a) $\kappa =1.75$; 
    (b) $\kappa =2.00$; (c) $\kappa =2.25$; (d) $\kappa =2.50$.
    The other parameters as in Fig. \ref{fig8}. The vertical black-solid line
    indicates the boundary between two different regimes of the decoupled CNLS
    equations, i.e., the unstable-stable regime for which $P_1 Q_{11} > 0$ and 
    $P_2 Q_{22} < 0$ ($k_2 < k_r$, left from the vertical line) and the 
    unstable-unstable regime for which $P_1 Q_{11} > 0$ and $P_2 Q_{22} > 0$
    ($k_2 > k_r$, right from the vertical line). 
    Note that $k_r$ depends weakly on the spectral index $\kappa$, but the 
    differences cannot be observed in this scale.
    }
    \label{fig8.2}
\end{figure}

The results shown in Fig. \ref{fig8} have been obtained with the same parameters 
as those used in Figs. \ref{fig6} and \ref{fig7}, except that the value of $k_1$ 
is now $k_1 =1.6 > k_r (\kappa)$. That means that the decoupled CNLS equations 
are either in the unstable-stable regime ($P_1 Q_{11} > 0$ and $P_2 Q_{22} < 0$) 
for $k_2 < k_r (\kappa)$, i.e., to the left from the black-solid vertical line,
while they are in the unstable-unstable regime 
($P_1 Q_{11} > 0$ and $P_2 Q_{22} > 0$) for $k_2 > k_r (\kappa)$, i.e., to the 
right of the black-solid vertical line. 
 In this case, we observe a change in the
obtained instability patterns, especially for $\kappa \geq 3$. For all values 
of $\kappa$, we again observe strong modulational instability with high values 
of $\Gamma$ for low $k_2$. For $\kappa =2$, modulational instability occurs in 
most parts of the areas of the $k_2 - K$ plane in which the decoupled CNLS 
equations are either in the unstable-stable or the unstable-unstable regime.    
For $\kappa \geq 3$ we again observe some shrinkage of the instability areas
with increasing $\kappa$. We can also observe for these values of $\kappa$ that 
the modulationally unstable areas are mostly in the unstable-stable regime than 
in the unstable-unstable regime of the decoupled CNLS equations.

In the case shown in Fig. \ref{fig8.2}, also, the patterns of the growth rate
$\Gamma$ on the $k_2 - K$ plane for a few values of $\kappa$ in the 
strongly nonthermal  region, i.e. for small values of $\kappa$. The evolution of 
these patterns is similar to the 
corresponding evolution described in Fig. \ref{fig8.2}. Thus, from Figs. 
\ref{fig6}-\ref{fig8.2} we can safely conclude that the most dramatic variation 
in the onset (region) of modulational instability on the $k_2 - K$ plane occurs 
in the interval of $\kappa$ values within the strongly nonthermal  region.

\begin{figure}[htp]
    \centering
    \includegraphics[width=12cm]{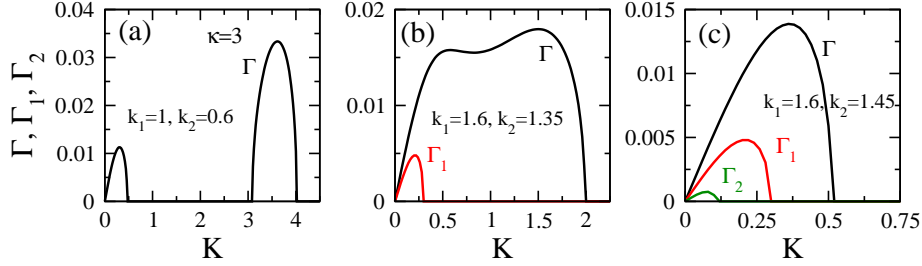}
    \caption{
    The growth rate $\Gamma$ of the CNLS system (black curves) and the growth 
    rates $\Gamma_1$ and $\Gamma_2$ (red and green curves, respectively) of the 
    first and the second decoupled CNLS Eqs. (\ref{eq_t05}) and (\ref{eq_t06} as 
    a function of the wavenumber of the perturbation $K$ for $\kappa =3$,
    $\Psi_{1,0} =\Psi_{2,0} =0.15$, and
    (a) $k_1 =1$ and $k_2 =0.6$, 
    for which $P_1 Q_{11} < 0$ and $P_2 Q_{22} < 0$ (stable-stable);
    (b) $k_1 =1.6$ and $k_2 =1.35$, 
    for which $P_1 Q_{11} > 0$ and $P_2 Q_{22} < 0$ (unstable-stable);
    (c) $k_1 =1.6$ and $k_2 =1.45$, 
    for which $P_1 Q_{11} > 0$ and $P_2 Q_{22} > 0$ (unstable-unstable).
    Note the difference in the horizontal scales.
    The growth rates $\Gamma_1$ and $\Gamma_2$ are always zero in (a), while
    $\Gamma_2$ is always zero in (b).
    }
    \label{fig9}
\end{figure}

\paragraph{Representative examples of growth rate variation.} 

It may be interesting to present some illustrative examples for the growth rate 
$\Gamma$ as a function of the perturbation wavenumber $K$ which may be distinctly 
different from those one may observe in a single NLS equation. In Fig. \ref{fig9}   
$\Gamma (K)$ curves are shown in three cases that differ in respect to the 
stability regime of the decoupled CNLS equations, for $\kappa =3$. Along with 
these curves, we also show the corresponding growth rates $\Gamma_1$ and 
$\Gamma_2$ obtained for the first and the second decoupled CNLS equation, 
respectively. In Fig. \ref{fig9}(a), in which $k_1 =1 < k_r (\kappa)$ and 
$k_2 =0.6 < k_r (\kappa)$, i.e., in the stable-stable regime of the decoupled 
CNLS equations, we obtain a $\Gamma (K)$ curve with two instability lobes. 
Interestingly, stronger modulational instability occurs in the second lobe,
for $K$ between $3$ and $4$, than in the first lobe for $K$ between $0$ and 
$0.5$ (low $K$). Since both the decoupled equations are modulationally stable, 
the corresponding growth rates $\Gamma_1$ and $\Gamma_2$ are both zero. 
In Fig. \ref{fig9}(b),
in which $k_1 =1.6 > k_r (\kappa)$ and $k_2 =1.35 > k_r (\kappa)$, i.e., in 
the unstable-stable regime of the decoupled CNLS equations, we obtain a 
$\Gamma (K)$ curve with two humps. Moreover, we obtain a non-zero 
$\Gamma_1 (\kappa)$ curve which has the commonly observed shape. 
A comparison between the two indicates that besides the difference in 
shape, the values of $\Gamma_1 (\kappa)$ are always significantly smaller 
than those of $\Gamma (\kappa)$. Moreover, the modulationally unstable 
interval in $K$ is much smaller in the first decoupled NLS equation than 
that of the coupled CNLS system. 
In Fig. \ref{fig9}(c), in which $k_1 =1.6 > k_r (\kappa)$ and 
$k_2 =1.45 > k_r (\kappa) =1.42$, i.e., in the unstable-unstable regime 
of the decoupled CNLS equations, we obtain three curves for the growth 
rates $\Gamma (K)$, $\Gamma_1$, and $\Gamma_2$ which have the same shape 
but very different size. As a result, the interval of the perturbation 
wavenumber $K$ in which the coupled CNLS system is modulationally unstable, 
is much larger that the corresponding one of either the first or the 
second decoupled NLS equation.

\section{Discussion and conclusions \label{section5}}

A pair of CNLS equations was derived for two co-propagating and interacting 
electrostatic wavepackets in a plasma fluid, comprising cold inertial ions 
evolving against a thermalized, kappa-distributed electron background. 
A multiple-scale technique has been adopted, similar to the well known Newell's 
method in nonlinear optics, for the reduction of the original plasma fluid model 
to the pair of CNLS equations which is generally non-integrable since the 
dispersion, nonlinearity, and coupling coefficients $P_j$ and $Q_{ij}$ 
($j=1,2$) depend on the carrier wavenumbers $k_1$ and $k_2$ (both assumed 
arbitrary) of the two interacting wavepackets. The exact mathematical 
expressions for those coefficients have been analytically obtained, and in 
fact depend parametrically on the plasma parameters too. Although $P_j$ for the
particular model considered here is negative, while it approaches asymptotically 
zero for large $k_j$, the coefficients $Q_{jj}$ can be either positive or 
negative. The value of $k_j$ that separates positive from negative $Q_{jj}$ 
values, $k_r$, depends on the value of the spectral index $\kappa$ which 
characterizes the kappa distribution of the electrons, i.e., 
$k_r =k_r (\kappa)$. The spectral index enters into the expressions of 
$Q_{jj}$ through the coefficients $c_1$, $c_2$, and $c_3$ of the Mc Laurin 
expansion of that distribution.
Interestingly, the $\kappa-$dependence of $k_r$ is not monotonous but it 
exhibits a minimum around $\kappa \simeq 2.5$ which has been set as a rough 
boundary between electrons far- and near-equilibrium by empirical 
investigations of Space plasmas. The $k_r (\kappa)$ dependence is weak for 
most values of $\kappa$, while it becomes stronger for $\kappa < 2$ where it 
increases abruptly until its divergence at $\kappa =3/2$. For very large 
values of $\kappa$, on the other hand, $k_r$ converges slowly to $1.47$ from 
below.

The coupling coefficients $Q_{12}$ and $Q_{21}$, which depend on both 
wavenumbers $k_1$ and $k_2$, may also be either positive or negative. While 
$P_j$ and $Q_{jj}$ have values of the order of unity, the values of $Q_{12}$ 
and $Q_{21}$ may also become very large positive or negative, implying very
strong coupling for the interacting electrostatic wavepackets in the plasma 
fluid. All this prescribes a multifaceted dynamical profile with strong 
interconnection among the various physical parameters.
A compatibility condition was derived through standard modulational 
instability analysis, in the form of a fourth degree polynomial in the 
frequency of the perturbation. By numerically calculating the roots of the 
compatibility condition and identifying the one with the largest imaginary 
part (growth rate), modulationally stable and unstable areas on the $k_2 - K$ 
plane have been identified. The instability growth rate $\Gamma$, mapped on 
that plane, reveals that modulational instability occurs in areas of the 
$k_2 - K$ plane (say, for fixed $k_1$), actually for all four stability 
regimes of the decoupled CNLS equation (i.e., those resulting by setting 
$Q_{12} =Q_{21} =0$), for all values of $\kappa$ used here.

Modulational instability in fact may occur even if both waves of the 
decoupled system are stable (stable-stable regime), since their strong 
nonlinear coupling may potentially destabilize the system in regions with 
large perturbation wavenumbers $K$. The instability that results from the 
nonlinear coupling between the two wavepackets is more intense in the case 
of equal amplitudes of the two waves (as those used in the present work); 
instability areas also occupy larger areas (``windows") on the $k_2 - K$ plane 
in the latter case. When one or both the waves of the decoupled system are 
unstable, their nonlinear coupling leads again to an unstable pair of 
wavepackets. A comparison of the growth rates of the two waves of the 
decoupled system with that of the waves in the coupled one reveals that the 
latter is always larger than the former; furthermore, the unstable areas on 
the $k_2 - K$ plane are larger than those of the waves in the decoupled system.
Also, as it becomes obvious from Figs. \ref{fig6}-\ref{fig8.2}, the most dramatic 
variation of the $\Gamma$ patterns on the $k_2 - K$ plane occurs for values of 
the spectral index $\kappa$ withing the strongly nonthermal  region, i.e., in the
interval $1.5 < \kappa < 2.5$. Moreover, as the numerical calculation of the 
growth rate $\Gamma$ indicate, the CNLS equations are modulationally unstable in 
almost all of the $k_2 - K$ plane shown in the figures below $\kappa \sim 1.9$.
For these values of $\kappa$, i.e., for $\kappa < 1.9$, the values of $\Gamma$
increase fast with decreasing $\kappa$ and diverge in the limit of 
$\kappa \rightarrow 1.5$.

Extensive numerical search support the conclusion that the nonlinear wave modes
of the CNLS equations are apparently always partly modulationally unstable, i.e., 
that there is an interval in the perturbation wavenumber $K$ for which 
modulational instability occurs. Although it cannot be considered as a definite 
proof our investigations shows that it is highly possible. As mentioned above, 
the instability is probably due to the large values of the coupling coefficients
$Q_{12}$ and $Q_{21}$ (as compared with the values of the coefficients $P_j$ and
$Q_{jj}$). Note that it is actually the product of the two, i.e., $Q_{12} Q_{21}$
which enters into $\Omega_c^4$ that appears in the right-hand-side of the 
compatibility condition Eq. (\ref{mi09}). However, it is possible to have 
modulationally stable nonlinear wave modes by ``eliminating'' $\Omega_c^4$ from
the right-hand-side of Eq. (\ref{mi09}). This is possible if we set one of the 
two coeffcients $Q_{12}$ and $Q_{21}$ equal to zero (in the considered model it 
is not possible to find $k_1$ and $k_2$ such as both these coefficients become 
zero). In fact, $Q_{12}$ is for example, zero along the curves shown in Fig.
\ref{fig4}(b) or as the contours shown in \ref{fig5} on the $k_1 - k_2$ plane. 
Along these curves $Q_{12} =0$ so that the CNLS equations are still at least 
partially coupled, but their perturbation fremquencies can be calculated from 
 Eq. (\ref{mi12}) which are the same as for the two decoupled CNLS equations.
 Thus, for modulational stability in this special case we can assert that 
 $P_1 Q_{11} < 0$ and $P_2 Q_{2} < 0$ simultaneously. This condition can be
 clearly satisfied within the considered model for $k_1, k_2 < k_r (\kappa)$.
 A bare look at Figs. \ref{fig4}(b) and \ref{fig5} is enough to see that there 
 are available $k_1$ and $k_2$ that satisfy the latter inequality.

The occurence of very broad modulational instability areas in the parameter 
space of $k_1$, $k_2$, and $K$, may on the other hand favor the emergence of 
vector solitons which may emerge in combinations of bright and dark components. 
The possible existence of such vector solitons, the determination of their 
parameters such as their width and amplitudes, as well as their stability, and 
how these are affected by the spectral index $\kappa$, will be the focus of 
future work.

\section{Acknowledgements}

Authors IK and NL gratefully acknowledge financial support from Khalifa University 
of Science and Technology, Abu Dhabi, United Arab Emirates, via the project 
CIRA-2021-064 (8474000412). IK also acknowledges financial support from KU via 
the project FSU-2021-012 (8474000352) as well as from KU Space and Planetary 
Science Center, Abu Dhabi,  United Arab Emirates,  via grant No. KU-SPSC-8474000336.

This work was completed during a long research visit by author IK to the 
National and Kapodistrian University of Athens, Greece. During the same period, 
IK also held an Adjunct Researcher status at the Hellenic Space Center, Greece. 
The hospitality of both hosts, represented by Professor D.J. Frantzeskakis and 
Professor I. Daglis, respectively, is warmly acknowledged.


\section{Declarations}

\subsection{Funding}

Authors IK and NL gratefully acknowledge financial support from Khalifa University 
of Science and Technology, Abu Dhabi, United Arab Emirates, via the project  
CIRA-2021-064 (8474000412) (PI Ioannis Kourakis). 
IK also acknowledges financial support from KU via the project 
FSU-2021-012 (8474000352) (PI Ioannis Kourakis) as well as from KU Space and 
Planetary Science Center, via grant No. KU-SPSC-8474000336 
(PI Mohamed Ramy Mohamed Elmaarry). 

\subsection{Competing Interests}

The authors have no relevant financial or non-financial interests to disclose.

\subsection{Author Contributions}

I. Kourakis contributed to the study conception, design, and methodology, 
while N. Lazarides contributed to the methodology, software development and 
numerics. Both authors contributed to the analysis of the results.  
The first draft was written by N. Lazarides  and the final manuscript 
by I. Kourakis. Both authors read and approved the final manuscript.

\subsection{Data Availability}

The datasets generated during and/or analysed during the current study are 
available from the corresponding author on reasonable request.




\begin{appendices}

\section{Coefficients - Analytical Expressions}\label{secA1}

\begin{eqnarray}
   C_{n,2,j}^{(2)} =\frac{(k_j^2 +c_1)}{6 k_j^2} 
                     \left[3 (4 k_j^2 +c_1) (k_j^2 +c_1) -2 c_2 \right],
\\
\label{eq131}
   C_{u,2,j}^{(2)} =\frac{\sqrt{k_j^2 +c_1}}{6 k_j^2} 
                     \left[3 (k_j^2 +c_1) (2 k_j^2 +c_1) -2 c_2 \right],
\\
   C_{\phi,2,j}^{(2)} =\frac{1}{6 k_j^2} \left[3 (k_j^2 +c_1)^2 -2 c_2 \right],
\end{eqnarray}

\begin{eqnarray}
   C_{n,2,\pm}^{(1)} =\frac{1}{D_\pm} \left\{ \left[ (k_1 \pm k_2)^2 +c_1 \right] 
        \left[ (k_1 \pm k_2) f_2^{(\pm)} +(\omega_1 \pm \omega_2) f_1^{(\pm)} \right] 
         +(k_1 \pm k_2)^2 f_3^{(\pm)}
   \right\},
   \nonumber \\
   \label{eq132}
   C_{u,2,\pm}^{(1)} =\frac{1}{D_\pm} \left\{ (k_1 \pm k_2) \left[ f_1^{(\pm)} 
    +(\omega_1 \pm \omega_2) f_3^{(\pm)} \right]
    +(\omega_1 \pm \omega_2) \left[ (k_1 \pm k_2)^2 +c_1 \right] f_2^{(\pm)}
   \right\},
   \\
   C_{\phi,2,\pm}^{(1)} =\frac{1}{D_\pm} \left\{ 
   (\omega_1 \pm \omega_2) \left[ f_1^{(\pm)} 
    +(\omega_1 \pm \omega_2) f_3^{(\pm)} \right] +(k_1 \pm k_2) f_2^{(\pm)} 
   \right\},
   \nonumber
\end{eqnarray}
where 
\begin{equation}
\label{eq133}
   D_\pm =(k_1 \pm k_2)^2 -(\omega_1 \pm \omega_2)^2 \left[ (k_1 \pm k_2)^2 +c_1 \right].
\end{equation}
The quantities $f_m^{(\pm)}$ ($m=1,2,3$) are defined in Eqs. (\ref{eq12.6})-(\ref{eq12.6}).

\begin{equation}
\label{eq134}
   C_{n,2,j}^{(0)} =\left( c_1 C_{\phi,2,j}^{(0)} +2 c_2 \right), \qquad
   C_{u,2,j}^{(0)} =\frac{1}{v_{g,j}} \left[ C_{\phi,2,j}^{(0)} +\frac{k_j^2}{\omega_j^2} \right], \qquad
   C_{\phi,2,j}^{(0)} =\rho_j \frac{1}{1 -c_1 v_{g,j}^2},
\end{equation}
where 
\begin{equation}
\label{eq135}
   \rho_j =-( k_j^2 +c_1 ) -2 c_1 +2 c_2 v_{g,j}^2. 
\end{equation}


\section{Right-Hand-Sides of the Perturbative Equations.}\label{secA2}%

\begin{eqnarray}
\label{eq12}
   {\cal F}_1 = -\frac{\partial (n_1)}{\partial t_1} 
                -\frac{\partial (u_1)}{\partial x_1} 
                -\frac{\partial (n_1 u_1)}{\partial x_0},
   \qquad
   {\cal F}_2 =-\frac{\partial u_1}{\partial t_1} 
               -\frac{\partial \phi_1}{\partial x_1} 
               -u_1 \frac{\partial u_1}{\partial x_0},
  \nonumber \\
   {\cal F}_3 =+c_2 \phi_1^2 -2 \frac{\partial^2 \phi_1}{\partial x_0 \partial x_1}.
\end{eqnarray}

\begin{eqnarray}
  \mu_{1,j}=
    -\left( \frac{k_j}{\omega_j} \right)^2 \frac{\partial \Psi_j}{\partial t_1}
    -\frac{k_j}{\omega_j} \frac{\partial \Psi_j}{\partial x_1},
\\
   \mu_{2,j}=
    -\frac{k_j}{\omega_j} \frac{\partial \Psi_j}{\partial t_1}
    -\frac{\partial \Psi_j}{\partial x_1}, 
\\
   \mu_{3,j}=
     -2 i k_j \frac{\partial\Psi_j}{\partial x_1}.
\nonumber
\end{eqnarray}

\begin{eqnarray}
   f_1^{(+)}=-(k_1 +k_2) \frac{k_1 k_2}{\omega_1 \omega_2} 
       \left( \frac{k_1}{\omega_1} +\frac{k_2}{\omega_2} \right),
   \\ 
   f_2^{(+)}=-(k_1 +k_2) \frac{k_1 k_2}{\omega_1 \omega_2},
   \\
   f_3^{(+)} =+2 c_2.
\end{eqnarray}

\begin{eqnarray}
   f_1^{(-)}=-(k_1 -k_2) \frac{k_1 k_2}{\omega_1 \omega_2} 
       \left( \frac{k_1}{\omega_1} +\frac{k_2}{\omega_2} \right),
   \\ 
   f_2^{(-)} =-(k_1 -k_2) \frac{k_1 k_2}{\omega_1 \omega_2},
   \\
  f_3^{(-)} =+2 c_2.
\end{eqnarray}

\begin{eqnarray}
\label{eq17}
  {\cal G}_1 =-\frac{\partial (n_2)}{\partial t_1} -\frac{\partial (u_2)}{\partial x_1}
   -\frac{\partial (n_1)}{\partial t_2} -\frac{\partial (u_1)}{\partial x_2}
   -\frac{\partial (n_1 u_2 +n_2 u_1)}{\partial x_0} -\frac{\partial (n_1 u_1)}{\partial x_1},
  \\
\label{eq18}
   {\cal G}_2 =-\frac{\partial u_2}{\partial t_1} -\frac{\partial \phi_2}{\partial x_1} 
   -\frac{\partial u_1}{\partial t_2} -\frac{\partial \phi_1}{\partial x_2}
   -\frac{\partial u_1 u_2}{\partial x_0} -u_1 \frac{\partial u_1}{\partial x_1},
   \\
\label{eq19} 
   {\cal G}_3 =-2 \frac{\partial^2 \phi_2}{\partial x_0 \partial x_1} 
   -2 \frac{\partial^2 \phi_1}{\partial x_0 \partial x_2}
   -\frac{\partial^2 \phi_1}{\partial x_1^2} +2 c_2 \phi_1 \phi_2 +c_3 \phi_1^3.
\end{eqnarray}

\begin{eqnarray}
\label{eq26}
  {\cal R}_{1,j} =-\left( \frac{\partial n_{2,j}^{(1)} }{\partial t_1} 
                  +\frac{\partial u_{2,j}^{(1)} }{\partial x_1} 
                         +\frac{\partial n_{1,j}^{(1)} }{\partial t_2} 
                         +\frac{\partial u_{1,j}^{(1)} }{\partial x_2}
  \right)
\nonumber \\
  -i k_j \left( n_{1,j}^{(-1)} u_{2,j}^{(2)} +n_{1,j}^{(1)} u_{2}^{(0)} 
     +n_{1,k_j}^{(-1)} u_{2,p}^{(1)} +n_{1,k_j}^{(1)} u_{2,m}^{(\ell_j)}
  \right)
  \nonumber \\
  -i k_j \left( u_{1,j}^{(-1)} n_{2,j}^{(2)} +u_{1,j}^{(1)} n_{2}^{(0)} 
     +u_{1,k_j}^{(-1)} n_{2,p}^{(1)} +u_{1,k_j}^{(1)} n_{2,m}^{(\ell_j)}
  \right),
  \\
  \label{eq27}
  {\cal R}_{2,j} =-\left( \frac{\partial u_{2,j}^{(1)} }{\partial t_1} 
                         +\frac{\partial \phi_{2,j}^{(1)} }{\partial x_1} 
                         +\frac{\partial u_{1,j}^{(1)} }{\partial t_2} 
                         +\frac{\partial \Psi_j }{\partial x_2}
                         \right)
\nonumber \\
  -i k_j \left( u_{1,j}^{(-1)} u_{2,j}^{(2)} +u_{1,j}^{(1)} u_{2}^{(0)} 
     +u_{1,k_j}^{(-1)} u_{2,p}^{(1)} +u_{1,k_j}^{(1)} u_{2,m}^{(\ell_j)}
  \right),  
  \\
   \label{eq28}
   {\cal R}_{3,j} =-2 i k_j \left( \frac{\partial \phi_{2,j}^{(1)} }{\partial x_1} 
   +\frac{\partial \Psi_j }{\partial x_2} \right)
\nonumber \\
   -\frac{\partial^2 \Psi_j }{\partial x_1^2}
   +2 c_2 \left( \Psi_j^\star \phi_{2,j}^{(2)} +\Psi_j \phi_{2}^{(0)} 
   +\Psi_{k_j}^\star \phi_{2,p}^{(1)} +\Psi_{k_j} \phi_{2,m}^{(\ell_j)} \right)
   \nonumber \\
   +3 c_3 \Psi_j \left( |\Psi_j|^2 +2 |\Psi_{k_j}|^2 \right),
\end{eqnarray}
where $j=1,2$, and $k_j$, $\ell_j$ defined as $\ell_1 =1$, $\ell_2 =-1$, 
$k_1 =2$, and $k_2 =1$.

\end{appendices}



\end{document}